\documentclass[
reprint,
nofootinbib,
amsmath,amssymb,
aps,
pra,
floatfix,
]{revtex4-2}

\usepackage{graphicx}
\usepackage{dcolumn}
\usepackage{bm}
\usepackage[colorlinks=true, allcolors= black]{hyperref}
\usepackage{soul}

\usepackage[most]{tcolorbox}
\usepackage[dvipsnames]{xcolor}
\usepackage{lipsum}
\usepackage{empheq}
\usepackage{csquotes}
\usepackage[b]{esvect}
\usepackage[caption=false]{subfig}
\usepackage[section]{placeins}
\usepackage{blindtext}

\newcommand{\green}[1]{\textcolor{Green}{#1}}
\newcommand{\iu}{{i\mkern1mu}}

\DeclareMathOperator{\sgn}{sgn} 

\makeatletter
\newcommand{\leftlabel}[1]{&&
  \refstepcounter{equation}\ltx@label{#1}%
  \tagform@{\theequation}&&}
\makeatletter

\makeatletter
\DeclareRobustCommand{\hvec}[1]{{\mathpalette\hvec@{#1}}}
\newcommand{\hvec@}[2]{%
  \vbox{\offinterlineskip
    \ialign{%
      \hfil##\hfil\cr
      $\m@th#1{}_{\rightharpoonup}$\kern-\scriptspace\cr
      $\m@th#1#2$\cr
    }%
  }%
}

\makeatletter
\renewcommand*\env@matrix[1][\arraystretch]{%
  \edef\arraystretch{#1}%
  \hskip -\arraycolsep
  \let\@ifnextchar\new@ifnextchar
  \array{*\c@MaxMatrixCols c}}

\begin{document}

\setstcolor{red}
\preprint{APS/123-QED}

\title{Analytic force-free jet from disk-fed rotating black holes}

\author{Luis Villarin}
\email{jlvillarin@nip.upd.edu.ph}
\author{Ian Vega}%
\email{ivega@nip.upd.edu.ph}
\affiliation{%
 National Institute of Physics, University of the Philippines, Diliman, Quezon City 1101, Philippines 
}%





\begin{abstract}
We present a new analytic model of a force-free electromagnetic jet launched from a disk-fed rotating black hole. The jet solution is force-free to first order in the black hole spin and is obtained through a systematic construction based on previously developed analytical methods. The black hole jet modeled here exhibits an asymptotically parabolic structure and is parametrized by the location of the current concentration and sign reversal on the thin magnetized disk in the equatorial plane. We find that the jet properties show negligible dependence on the disk parameter, suggesting a possible universality of slowly rotating black hole jets with respect to disk structure. Our jet model reproduces the key features expected from the Blandford-Znajek mechanism.
\end{abstract}


\maketitle

\section{Introduction}

Astrophysical jets associated with rotating black holes represent some of the most energetic phenomena observed in the universe. These jets are predominantly observed in active galactic nuclei (AGN), where a supermassive black hole, with mass in the range $ \sim 10^{5}-10^{10} M_{\odot}$, steadily accretes hot gas and energetic particles and persistently emits electromagnetic radiation through the jet across frequencies from radio to gamma rays~\cite{Heckman_2011,Alexander_2012,Webster_2002}. For instance, estimates of the jet power of $\mathrm{M87}^{*}$ from various measurements range from $10^{42}$ to $10^{43}\,\mathrm{erg} \cdot \mathrm{s^{-1}}$~\cite{refId0,Jianchao_2017}, about a billion times more luminous than the Sun. Also, jets are observed across all black hole mass scales, including stellar-mass black holes, which typically host accretion disks formed via mass transfer from their companion stars or in the aftermath of binary neutron star mergers~\cite{Shakura:1972te,Narayan:1992iy}. 

The black hole jets are widely believed to be driven by large-scale, strong magnetic fields threading the black hole horizon, consistent with observational evidence~\cite{Akiyama_2021, Goddi_2021}. The accretion disk surrounding the black hole serves as the source of a strong magnetic field, as the infalling gas is thought to become a magnetized plasma due to viscous heating from friction and differential rotation, while dynamical instabilities and turbulence in the flow amplify the generated magnetic field~\cite{Abramowicz:2011xu,Frank_King_Raine_2002}. Numerical simulations have strengthened this view that the relativistic jets arise from poloidal magnetic fields generated by accretion flows onto black holes~\cite{Liska_2020, Tchekhovskoy_2011,Kim:2025}. However, the underlying physics of how these strong magnetic fields extract energy from black holes remains an open question.

The Blandford-Znajek (hereafter, BZ) mechanism~\cite{BlandfordZnajek1977} is the prevailing theory for energy extraction from rotating black holes. BZ showed that when the plasma's stress-energy tensor is negligible compared to that of the electromagnetic field in a stationary, axisymmetric black hole magnetosphere, the black hole’s rotational energy can be extracted by the magnetic field threading the horizon, in a manner analogous to a Penrose process~\cite{Komissarov:2008yh}. In this regime, where plasma inertia and pressure are negligible relative to the electromagnetic energy and stresses, the magnetosphere is well described by force-free electrodynamics (FFE). Within this framework, the energy extraction process is determined entirely by the structure of the magnetic field, independent of the plasma microphysics and detailed magnetohydrodynamic processes. Also, while FFE was first introduced in a pulsar magnetosphere model~\cite{Goldreich:1969} and has since been widely applied in other astrophysical contexts, its application to rotating black holes is distinguished by the requirement that the electromagnetic fields satisfy a universal regularity condition at the horizon~\cite{Znajek:1977}.

While numerical simulations are often more accessible and insightful, analytical approaches to black hole force-free electrodynamics remain indispensable for enforcing regularity conditions and ensuring global consistency of the solutions. However, the equations governing FFE in Kerr spacetime, which describes astrophysical black holes in general relativity, are highly nonlinear and appear to be analytically intractable, even under the assumptions of stationarity and axisymmetry, and physically relevant analytic solutions remain scarce. Few FFE solutions in the Kerr background are known~\cite{Adhikari_etal_2024, Brennan_2013, Menon:2015, Compere_2016}, with only a handful describing force-free jets from disk-accreting rotating black holes~\cite{BlandfordZnajek1977, Beskin:2010iba}. These black hole jet solutions are obtained to first order in the dimensionless black hole spin parameter using the well-known Blandford–Znajek perturbation method~\cite{BlandfordZnajek1977}, in which the electromagnetic properties are expressed as power series in the spin parameter, which serves as the perturbation parameter. At first-order in this perturbation, the magnetic flux function corresponds to the flux function of a known vacuum solution in Schwarzschild spacetime, while the angular velocity of the magnetic field and the current are linear in the angular velocity of the horizon. In the model of the black hole hyperbolic jet, a vacuum solution in flat spacetime was first derived and promoted to Schwarzschild spacetime via a unique canonical mode-mode mapping before applying the BZ perturbation method~\cite{Gralla_etal_2016}. However, higher-order perturbative expansions require careful treatment because inconsistencies may arise~\cite{Tanabe:2008,Grignani:2018, Grignani_2020}, which can nevertheless be resolved using the method of matched asymptotic expansions~\cite{Armas_2020,Camilloni:2022}.

Efforts to solve the FFE equations have increasingly leaned toward geometrical approaches. Recent developments have highlighted the importance of the underlying geometric structure of the background spacetime of force-free electromagnetic fields that opened new avenues for constructing analytic FFE solutions. Building on the covariant formulation of FFE introduced by Uchida~\cite{Uchida_1997,Uchida:1997b}, Gralla and Jacobson~\cite{Gralla_Jacobson:2014} developed a fully spacetime-covariant approach to FFE using exterior calculus. With this formulation, they were able to recover several known FFE solutions, as well as the properties of black hole force-free magnetospheres, in a more concise and elegant manner. Also, in this formalism, force-free fields naturally appear as so-called degenerate fields, allowing them to be associated with two-dimensional foliations of the background spacetime. These foliations represent the time evolution of magnetic field lines of force-free magnetospheres, to which the force-free plasma is confined. 

In addition, Menon~\cite{Menon_2021} introduced a geometric approach to FFE in terms of spacetime-foliating inertial tetrads, or inertial frames, on the background spacetime. Any inertial tetrad whose two basis vectors span an involutive distribution foliates the spacetime. In Ref.~\cite{Menon_2021}, he showed that a foliation, and thus the tetrad, associated with a force-free field satisfies a specific geometric condition, and the tetrad uniquely determines the force-free field. Hence, one passes from the field to the tetrad as the fundamental variable in this approach. Also, he showed that some known tetrads on Kerr spacetime obey the geometric condition and correspond to previously known force-free fields. Moreover, Adhikari \textit{et al.}~\cite{Adhikari_etal_2024} noted that inertial frames are related by Lorentz transformations and exploited this fact to develop a technique for finding force-free solutions, through which they obtained new classes of force-free solutions around Kerr black holes. In general, however, their technique ,may present analytic difficulties comparable to those encountered in directly solving the highly nonlinear force-free equations in Kerr spacetime. Nevertheless, it proves especially effective in flat spacetime, as demonstrated in this paper.

In this paper, we build a novel analytic force-free jet model around a disk-fed rotating black hole. The model is obtained through a systematic construction of force-free solution rather than a derivation from first principles. We combine previous methods to arrive at our jet model: deriving a new vacuum solution in flat spacetime using a procedure similar to the technique of Adhikari \textit{et al.}~\cite{Adhikari_etal_2024}, applied to the tetrad of hyperbolic solution, followed by promoting the solution to Schwarzschild background by the unique canonical mapping introduced in Ref.~\cite{Gralla_etal_2016}, and finally applying the BZ perturbation method of Ref.~\cite{BlandfordZnajek1977}. More specifically, we derive new vacuum degenerate fields in flat spacetime sourced by a thin magnetized disk. Among these new fields, we identify a physically reasonable solution, parameterized by the location of sign reversal and concentration of current on the disk, that exhibits asymptotically parabolic magnetic field lines while avoiding pathological current divergences. We then promote this solution to a Schwarzschild background by first expanding it near the origin and at infinity, and subsequently applying the canonical mode-mode mapping of Ref.~\cite{Gralla_etal_2016}. Then, we use the promoted solution as a seed for the BZ perturbation method to find a slowly rotating solution to first order in the black hole spin. Afterwards, we impose the horizon regularity condition together with a well-established asymptotic condition of the magnetic field manifested by the solution to determine the angular velocity and current of the jet. 

In addition, this work provides the first application of the approach of Ref.~\cite{Adhikari_etal_2024} in flat spacetime, revealing our solutions that are not found by other methods, such as in Ref.~\cite{Compere_2016}. Our procedure, on the other hand, differs from Ref.~\cite{Adhikari_etal_2024} in that we derive the tetrad of a known field, specifically the vacuum hyperbolic solution, rather than starting from a known tetrad. Also, in conjunction with other methods, it offers an alternative approach for constructing analytic jets around pulsars and black holes.

We organized this paper as follows. In Sec.~\ref{sec:II}, we briefly review force-free electrodynamics from a geometric perspective, focusing on the degeneracy of force-free fields and their spacetime foliations. We also discuss in this section exact vacuum solutions in flat spacetime that are useful for constructing our jet solution. In Sec.~\ref{sec:III}, we derive new vacuum solutions from the hyperbolic solution using the geometric methods mentioned earlier. In Sec.~\ref{sec:IV}, we promote the physically viable solution to Schwarzschild spacetime and perform the BZ perturbation method. In the same section, we determine the angular velocity and current associated with the black hole jet and analyze these properties, as well as other important features of the jet, specifically its total electromagnetic energy flux and effective resistance. Readers only interested in the black hole jet may skip the preceding sections and proceed directly to Sec.~\ref{sec:IV}. Our metric signature is always $(-,+,+,+)$ and we work in Heaviside-Lorentz and geometrized units ($G = c =  \mu_{0}=1$). We adopt the orientation $dt \wedge dz \wedge d\rho \wedge d\varphi$ in flat spacetime with cylindrical coordinates $(t,z,\rho,\varphi)$.

\section{Theory of force-free fields}\label{sec:II}

In the first two parts of this section, we present a brief, self-contained review of the covariant spacetime formulation and the geometric structure of force-free electrodynamics, following Refs.~\cite{Gralla_Jacobson:2014}~and~\cite{Menon_2021}. We conclude this section by discussing the mapping of vacuum fields from flat to Schwarzschild spacetime, which is relevant for the analytic derivation of our analytic jet model.

\subsection{Force-free electrodynamics}\label{sec:IIA}

In a force-free setting, the electromagnetic four-force density is zero, 
\begin{equation}\label{eq:2.1}
    F_{ab}j^{b} = 0 \, , 
\end{equation}
where $F_{ab}$ is the Faraday $2$-form and $j^{a}$ is the four-current. Together with Maxwell's equations, 
\begin{align}
\nabla_{[a}F_{bc]} &= 0 \, , \label{eq:2.2} \\
\nabla_{b} F^{ab} &= j^{a} \, , \label{eq:2.3} 
\end{align}
force-free electrodynamics can be expressed as 
\begin{equation}
    \nabla_{[a}F_{bc]} = 0 \, , \:\:\: F_{ab}\nabla_{c}F^{bc} = 0 \, . \label{eq:2.4 }
\end{equation}

In the language of differential forms, Maxwell's equations are written as 
\begin{eqnarray}
dF = 0 \, , \label{eq:2.5} \\
d*F = J \, , \label{eq:2.6} 
\end{eqnarray}
where the star `$*$' denotes Hodge dual with respect to the background spacetime\footnote{The Faraday tensor is written as $F = \dfrac{1}{2} \,F_{ab} dx^{a} \wedge dx^{b}$, and the current-charge three-form $J$ is related to the four-current via $j^{a} = \dfrac{1}{6}\epsilon^{abcd}J_{bcd}$.}. The homogeneous Maxwell equations~\eqref{eq:2.5} state that the Faraday $2$-form is closed. Although not all closed forms are globally exact, we may `locally' write $F = dA$ for some potential one-form $A$. In force-free electrodynamics, however, the field admits a more constrained structure. 

Force-free fields belong to the class of degenerate fields that obey the condition
\begin{equation}
    F_{[ab}F_{cd]} = 0 \, , \label{eq:2.7}
\end{equation}
This condition implies that a degenerate field is a wedge product of two one-forms,
\begin{equation}
    F = \alpha \wedge \beta \,. \label{eq:2.8}
\end{equation}
The degeneracy of force-free fields can be realized by rewriting the force-free condition as
\begin{equation}
    F_{[ab}F_{cd]}j^{d} = 0  \:. \label{eq:2.9}
\end{equation}
Also, this condition amounts to the vanishing of the Lorentz scalar $\vv{E} \cdot \vv{B}$ in a local frame. It is also worth noting that not all degenerate fields are force-free, whereas all force-free fields are degenerate. 

From the degeneracy condition and the homogeneous Maxwell equations, a force-free field is generally given by
\begin{equation}
    F = d\phi_{1} \wedge d\phi_{2} \:, \label{eq:2.10}
\end{equation}
where $\phi_{1}$,$\phi_{2}$ are functions called the \textit{Euler potentials}. We may rephrase the force-free condition as 
\begin{equation}
    F_{a[b}J_{cde]} = 0 . \label{eq:2.11}
\end{equation}
By Eqs.~\eqref{eq:2.10} and~\eqref{eq:2.11}, the Euler potentials of a force-free field should then satisfy the equations,
\begin{equation}
    d\phi_{1} \wedge J = 0 = d\phi_{2} \wedge J \,. \label{eq:2.12}
\end{equation}
The remaining equations to be satisfied by the field are the inhomogeneous equations that finally provide the complete description of the force-free electrodynamics,
\begin{equation}
    d\phi_{i} \wedge d * (d\phi_{1} \wedge d \phi_{2}) = 0 \:,\:\:\: i = 1,2 \:.  \label{eq:2.13}
\end{equation}
Obviously, a vacuum degenerate Maxwell field automatically satisfies Eq.~\eqref{eq:2.13}, hence it is trivially force-free. In what follows, we refer to vacuum degenerate Maxwell field simply as vacuum degenerate field, or vacuum field. 

\subsection{Geometric structure}\label{sec:IIB}

The Euler potentials determine the nature of the force-free field. By taking $(d\phi_{1})_{a}$ and $(d\phi_{2})_{a}$ to be orthogonal, the square of the magnitude of the field strength is 
\begin{equation}
    F^{2} = F_{ab}F^{ab}= 2(B^{2}-E^{2}) = 2 \|d\phi_{1}\|^2 \|d\phi_{2}\|^{2} \:, \label{eq:2.14}
\end{equation}
where $\| \alpha \|^{2} = g^{ab}\alpha_{a}\alpha_{b}$ is the squared-norm. Since there are no two orthogonal timelike vectors, the field is magnetically dominated ($F^{2}>0$) if and only if both vectors $(d\phi_{1})^{a}$ and $(d\phi_{2})^{a}$ are spacelike. Such a field describes a force-free magnetosphere, in which there always exists a Lorentz frame where the electric field vanishes. Accordingly, we restrict attention to magnetically-dominated force-free field from this point forward.

The degeneracy of a force-free field has profound implications on the geometric structure of the background spacetime. At each point, there exists a two-dimensional vector space that annihilates the force-free field, namely the kernel of $F$, denoted as ker $F$. A kernel of $F$ is orthogonal to the space spanned by vectors $(d\phi_{1})^{a}$ and $(d\phi_{2})^{a}$, and forms an involutive distribution. By Frobenius theorem, it follows that kernels of $F$ are tangent to two-dimensional integral submanifolds. These integral surfaces are foliations of the spacetime, and are called \textit{field sheets} by Gralla and Jacobson~\cite{Gralla_Jacobson:2014}, and \textit{flux surfaces} by Uchida~\cite{Uchida_1997}. In force-free magnetospheres, these submanifolds are Lorentzian and describe the spacetime evolution of the magnetic field lines. Also, field sheets are a general feature of degenerate fields.

Menon~\cite{Menon_2021} established a geometric correspondence between field sheets and local inertial frames on a spacetime manifold $M$. At each point $p \in M$, there exists a local inertial frame $(e_{0},e_{1},e_{2},e_{3})$ defined on an open neighborhood $U_{p}$ of $p$ such that the timelike basis $e_{0}$ and spacelike basis $e_{1}$ span the kernel of a degenerate field $F$ tangent to the field sheet restricted to $U_{p}$. The mean curvature field on the field sheet can be expressed as
\begin{equation}
    H = \dfrac{1}{2} \biggl[ - \Pi(e_{0},e_{0}) + \Pi(e_{1},e_{1}) \biggr] \, ,
    \label{eq:2.15}
\end{equation}
where $\Pi$ is the extrinsic curvature or the \textit{second fundamental form} of the sheet that takes tangent vector fields, say $V$ and $W$, and gives 
\begin{equation}
    \Pi(V,W) = (\nabla_{V}W)^{\perp} \,.
    \label{eq:2.16}
\end{equation}
The symbol $\perp$ denotes the projection operator that takes the normal component of the vector on the field sheet. We can also define a `dual' mean curvature, even if the complementary orthogonal space spanned by $e_{2}$ and $e_{3}$ does not form an involutive distribution, as
\begin{equation}
    \Tilde{H} = \frac{1}{2} \biggl[ \Pi(e_{2},e_{2}) + \Pi(e_{3},e_{3}) \biggr] \,. \label{eq:2.17}
\end{equation}

Menon~\cite{Menon_2021} proved a remarkable theorem, stating that, up to a constant in $h$, a degenerate field expressed as
\begin{equation}
    F = h \, e^{\flat}_{2} \wedge e^{\flat}_{3} \,,
    \label{eq:2.18}
\end{equation}
with the dual field
\begin{equation}
    *F = h \, e^{\flat}_{0} \wedge e^{\flat}_{1} \,,
    \label{eq:2.19}
\end{equation}
is the unique magnetically-dominated force-free field, if and only if, 
\begin{equation}
    dH^{\flat} + d\Tilde{H}^{\flat} = 0 \, , \label{eq:2.20}
\end{equation}
where `$\flat$' denotes musical isomorphism. As a corollary, 
\begin{equation}
    d(\ln{h}) = 2(H^{\flat} + \Tilde{H}^{\flat}) \,. \label{eq:2.21}
\end{equation}
Eqs.~\eqref{eq:2.20} and~\eqref{eq:2.21} are statements that $H^{\flat} + \Tilde{H}^{\flat}$ should be closed and exact. Also, if the distribution spanned by $e_{2}$ and $e_{3}$ are also involutive, the field is vacuum degenerate. 

\subsection{Vacuum degenerate fields}\label{sec:IIC}

Although vacuum degenerate fields are trivial solutions of the force-free equations, they provide a natural starting point for perturbative constructions of solutions in Kerr spacetime, such as the Blandford-Znajek perturbation method. Gralla, Lupsasca, and Rodriguez~\cite{Gralla_etal_2016} derived a first-order jet solution in the black hole spin $a$ by extending a known vacuum solution in flat spacetime to Schwarzschild spacetime and performing the BZ perturbation method. 

In the following sections, we adopt a similar procedure after we identify an interesting vacuum field in flat spacetime using the geometric approaches. Here, we briefly discuss the canonical mapping of vacuum degenerate solutions in flat spacetime to solutions in Schwarzschild spacetime. Also, we work with stationary, axisymmetric fields, which are relevant for the discussions of energy extraction from a rotating black hole. This section mainly revisits Appendix B of Ref.~\cite{Gralla_etal_2016}.

A stationary, axisymmetric degenerate field in a background spacetime that possesses the same symmetries is fully characterized by three functions: the poloidal magnetic flux function $\psi(r,\theta)$, angular velocity of the poloidal magnetic field lines $\Omega_{F}(\psi)$, and total poloidal electric current $I(r,\theta)$\footnote{We refer the reader to Sec.~7 of Ref.~\cite{Gralla_Jacobson:2014} for a detailed discussion of the poloidal and toroidal decomposition of stationary, axisymmetric spacetimes, as well as the general stationary, axisymmetric fields.}. The flux function and the poloidal magnetic field lines uniquely determine one another, and the latter is the intersection of the poloidal surface with the field sheet. Thus, the flux function can be taken to depend only on the level set function of a poloidal curve, represented by $u(r,\theta
)$ for instance. Under level-set reparametrization, $u \rightarrow f(u)$, the force-free equation is covariant and the field remains unchanged (see Ref.~\cite{Compere_2016}). Only a handful of classical curves are known to represent force-free magnetospheres around black holes. 

In terms of the three functions, the field may always be taken to have the form 
\begin{align}\label{eq:2.22}
    2\pi F &= d\psi \wedge \eta + I \sqrt{-\dfrac{g^{P}}{g^{T}}} \, dr \wedge d\theta \, , \notag \\
    \eta &\equiv d\varphi - \Omega_{F} \, dt, 
\end{align}
where $g^{T}$ is the metric determinant of the toroidal submanifold generated by the time-translation and axial rotation Killing vectors, $\partial_{t}$ and $\partial_{\varphi}$, respectively, and $g^{P}$ is the metric determinant of the poloidal surface orthogonal to the Killing vectors with coordinates $(r,
\theta)$\footnote{The corresponding Euler potentials are $\phi_{1}= \psi(r,\theta)$ and $\phi_{2} = \psi_{2}(r,\theta) + \varphi - \Omega_{F}(\psi) t$. The function $I$ appears on the Faraday two-form by redefinition of $\psi_{2}$. Also, we use the convention from Ref.~\cite{Gralla_etal_2016}.}. This field must then satisfy the force-free equations in Eq.~\eqref{eq:2.13} to be force-free. One of the force-free equations immediately implies $I = I (\psi)$, from which it follows that the current flows along the magnetic field lines. Meanwhile, the other equation is a second-order nonlinear partial differential equation relating $\psi$, $\Omega_{F}(\psi)$, and $I(\psi)$, known as the \textit{stream equation} or the \textit{Grad-Shafranov equation}. The explicit form of the full stream equation in Kerr spacetime, expressed in Boyer-Lindquist coordinates, is
\begin{align}\label{eq:2.23}
    &\eta_{\alpha} [\partial_{r}(\eta^{\alpha} \Delta \sin{\theta} \partial_{r} \psi ) + \partial_{\theta} (\eta^{\alpha} \sin{\theta} \partial_{\theta} \psi)] + \dfrac{\Sigma II' }{\Delta \sin{\theta}} = 0, \notag \\ 
    &\Sigma \equiv r^{2}+a^{2}\cos^{2}{\theta} , \:\: \Delta \equiv r^{2}-2Mr+a^{2} ,
\end{align}
where overprime denotes differentiation with respect to $\psi$ and $M$ is the black hole mass, and has been presented extensively in the literature (see, e.g., Refs.~\cite{Camilloni:2022,Beskin:1997}). Its counterparts in flat and Schwarzschild spacetimes are presented, for instance, in Refs.~\cite{Beskin:2010iba,Gralla_etal_2016,Beskin:1997, Ghosh1999TheSO}.

In the vacuum case ($I = \Omega_{F} = 0$), a stationary, axisymmetric field is then given by
\begin{equation}\label{eq:2.24}
    2\pi F = d \psi(r,\theta) \wedge d \varphi \, . 
\end{equation}
The linear superposition of such fields remains degenerate and stationary-axisymmetric.

Meanwhile, the stream equation in the vacuum case in flat spacetime reduces to
\begin{equation}\label{eq:2.25}
    r^{2} \partial^{2}_{r} \psi + \partial^{2}_{\theta} \psi - \cot{\theta} \, \partial_{\theta} \psi = 0  \, , 
\end{equation}
whereas in the Schwarzschild spacetime it becomes
\begin{equation}\label{eq:2.26}
    r(r-2M) \partial^{2}_{r} \psi +2M \partial_{r} \psi+ \partial^{2}_{\theta} \psi - \cot{\theta} \, \partial_{\theta} \psi = 0  \, . 
\end{equation}

We can make an identification between the solutions of Eqs.~\eqref{eq:2.25} and~\eqref{eq:2.26}, since they share the same angular equation and agree entirely in the limit $ r \rightarrow \infty$ or $M \rightarrow 0$, upon identifying the flat spacetime coordinates with the Schwarzchild coordinates. By writing the flux function as $\psi_{l} \propto R_{l} (r)\Theta_{l}(\theta)$, Eq.~\eqref{eq:2.26} separates into
\begin{flalign}
    &\frac{d}{d\theta} \biggl( \frac{1}{\sin{\theta}} \frac{d\Theta_{l}}{d\theta} \biggr) + \frac{l (l+1)}{\sin{\theta}} \Theta_{l} = 0 \, , \label{eq:2.27} \\[1em]
    &\frac{d}{dr}\biggl[ \biggl( 1- \frac{2M}{r} \biggr) \frac{dR_{l}}{dr}\biggr] - \frac{l(l+1)}{r^{2}}R_{l} = 0 \,,  \label{eq:2.28}
\end{flalign}
where $l(l+1) >0$ serves as a separation constant. 

The angular equation~\eqref{eq:2.27} is of Sturm-Liouville form, and its eigenfunctions can be classified into odd modes with $l = 2k-1$ and even modes with $l = 2k$, where $k$ is a positive integer. Hence, the angular harmonics form a complete set of basis eigenfunctions that is mutually orthogonal with respect to the weight $\csc{\theta d\theta}$. The eigenfunctions that vanish at the poles are given by the following\green{~\cite{Gralla_etal_2016}}:
\begin{align}
    \Theta_{2k-1}(\theta) &= \phantom{}_{2}F_{1}[ k -1/2, -k , 1/2, \cos^{2}{\theta}] \, , 
    \label{eq:2.29} \\[0.5em]
    \Theta_{2k}(\theta) &= \phantom{}_{2}F_{1}[ k + 1/2 , - k, 3/2, \cos^{2}{\theta}] \, \cos{\theta} \,. \label{eq:2.30}
\end{align}
The $l=0$ angular solution is excluded from the set of the eigenfunctions since it is not orthogonal to all of them. 

Meanwhile, the radial equation in flat spacetime has two kind of solutions: solutions that are regular at the origin (but not at infinity), $r^{l+1}$, and regular at infinity (but not at the origin), $r^{-l}$. Analogous radial solutions exist in Schwarzschild spacetime that are regular only at the horizon, $R^{<}_{l}$, or at infinity, $R^{>}_{l}$. These can be normalized such that, in the limit $r \rightarrow \infty$ or $M \rightarrow 0$, they reduce to the corresponding flat spacetime solutions, i.e., $R^{<}_{l} \rightarrow r^{l+1}$ and $R^{>}_{l} \rightarrow r^{-l}$. These radial solutions are presented in Refs.~\cite{Gralla_etal_2016,Ghosh1999TheSO}, and here we quote $R^{<}_{l}$ and provide an algebraic simplification for $R^{>}_{l}$, as follows
\begin{align}
    R^{<}_{l}(r) &= \frac{r^{2}}{2} (-2M)^{l-1} \frac{\Gamma(l+2)^{2}}{\Gamma(2l+1)} \notag \\
    &\times \, \phantom{}_{2}F_{1} \biggl[ l+2, 1-l ; 3 ; \frac{r}{2M} \biggr] \, , \label{eq:2.31} \\
 R^{>}_{l}(r) &= 
    \frac{2}{\sqrt{\pi}} \biggl( \dfrac{r}{4} \biggr)^{-l} \frac{\Gamma(l+\frac{3}{2})}{(l+1)\Gamma(l)}  \notag \\
    &\times \sum^{\infty}_{j = 0} \frac{\Gamma(l+j)\Gamma(l+j+2)}{\Gamma(2l+2+j)\Gamma(j+1)} \biggl( \dfrac{2M}{r}\biggr)^{j} \label{eq:2.32}\\
    &= r^{-l} \phantom{}_{2} F_{1} \biggl[l,l+2, 2l+2, \dfrac{2M}{r} \biggr] .
    \label{eq:2.33}
\end{align}
We obtained the most compact analytic expression to $R^{>}_{l}(r)$, provided in~Eq.~\eqref{eq:2.33}, by expanding the expression inside the braces of Eq.~B.12 in Ref.~\cite{Gralla_etal_2016} about $2M/r = 0$ and resumming the resulting series.

\section{New Vacuum fields in flat spacetime}\label{sec:III}
Any other local inertial frame can be generated from a known tetrad $(e_{0} , e_{1} , e_{2} , e_{3})$ on a background spacetime via a spacetime-dependent homogeneous Lorentz transformation $\Lambda(x)$, 
\begin{equation}\label{eq:2.34}
\begin{pmatrix}
    \Bar{e}_{0} \\
    \Bar{e}_{1} \\
    \Bar{e}_{2} \\
    \Bar{e}_{3}
\end{pmatrix}
= \Lambda(x) 
\begin{pmatrix}
    e_{0} \\
    e_{1} \\
    e_{2} \\
    e_{3}
\end{pmatrix} \, ,
\end{equation}
where $\Lambda$ satisfies $\eta = \Lambda^{T} \eta \Lambda$. Given a known inertial frame or a frame associated with a force-free solution, one may use the theorem discussed in Sec.~\ref{sec:II} to construct a new force-free field via an appropriate Lorentz transformation. In general, this procedure does not guarantee the existence of a new physically meaningful solution, since one must still provide a transformed tetrad that follows the conditions in~\eqref{eq:2.20}~and~\eqref{eq:2.21}, while imposing regularity conditions separately. Nevertheless, it provides a systematic method for the search for new exact solutions in arbitrary background spacetimes. This is the technique first introduced by Adhikari \textit{et al.}~\cite{Adhikari_etal_2024} in Kerr spacetime, where they succeeded in finding new exact solutions.

In flat spacetime, the only transformation that can generate a new vacuum degenerate field from the tetrad of a known one is an orthogonal transformation in the poloidal plane. To the best of our knowledge, an explicit proof of this simple statement has not been presented in the literature, and we therefore provide one in Appendix~\ref{apx:A}. But, in this work, we only consider the improper rotation
\begin{equation}\label{eq:3.35}
\Lambda(x) = 
\begin{pmatrix}[2]
    1 & 0 & 0 & 0 \\
    0 & \dfrac{1}{\sqrt{1+L^{2}}} & \dfrac{L}{\sqrt{1+L^{2}}} & 0 \\
    0 & \dfrac{L}{\sqrt{1+L^{2}}} & - \dfrac{1}{\sqrt{1+L^{2}}}  & 0 \\
    0 & 0 & 0 & 1
\end{pmatrix}    \, ,  
\end{equation}
where $L$ is a function of the poloidal coordinates. 

In this section, we apply the previous method to the vacuum hyperbolic field in flat spacetime, sourced by a finite disk around a compact object, to obtain new solutions. The hyperbolic field is parametrized by an inner disk radius, but this parameter acquires a different physical interpretation in the new solutions constructed here.  

\subsection{Parametrized solutions}\label{sec:IIIA}
Among known vacuum fields in flat spacetime, the hyperbolic field is unique in being parametrized by the inner radius of a thin equatorial disk surrounding a compact object; for this reason, we choose it as our seed solution. The solution can alternatively be obtained using the foliation-dependent method presented in Ref.~\cite{Compere_2016}, in which an output of a function builder describing possible magnetic field lines of a force-free field is checked to determine whether it satisfies the universal `foliation condition', expressed in terms of the representative $u(r,\theta)$. In principle, we can arrive at the new solutions presented here using this method, but we later find that the solutions are expressed in a relatively more complicated expression of poloidal coordinates that the function builder used in Ref.~\cite{Compere_2016} cannot capture.

The level set function on which the magnetic flux of hyperbolic field depends, in cylindrical coordinates, is
\begin{equation}\label{eq:3.36}
    u = \frac{\sqrt{(\rho +d)^{2}+z^{2}} - \sqrt{(\rho -d)^{2} +z^{2}}}{2d} \, ,
\end{equation}
where $d$ is the inner disk radius. The flux function of the hyperbolic field in the vacuum case is~\cite{Beskin:2010iba, Gralla_etal_2016}
\begin{equation}\label{eq:3.37}
    \psi(u) = \psi_{0} (1-\sqrt{1-u^{2}}) \,,
\end{equation}
where $\psi_{0}$ is a constant factor\footnote{The hyperbolic solution has an exact rotating counterpart in flat spacetime~\cite{Gralla_etal_2016,Compere_2016}.}. The radial and vertical fields arise as special cases of the hyperbolic field in the limits $d \rightarrow 0$ and $d \rightarrow \infty$, respectively. The tetrad associated with the vacuum hyperbolic field can be easily derived using Eqs.~\eqref{eq:2.18} \&~\eqref{eq:2.19} and the orthogonality condition of tetrad basis, and is given by
\begin{equation}\label{eq:3.38}
\begin{pmatrix}
    e_{0} \\
    \\
    e_{1} \\
    \\
    e_{2} \\
    \\
    e_{3}
\end{pmatrix}
= 
\begin{pmatrix}
    \partial_{t} \\
    \rho \sqrt{\dfrac{1-u^{2}}{\rho^{2} - d^{2} u^{4}}} \, \partial_z + u \sqrt{\dfrac{\rho^{2} -d^{2}u^{2} }{\rho^{2} -d^{2}u^{4}}} \, \partial_\rho \\
    u \sqrt{\dfrac{\rho^{2} -d^{2}u^{2} }{\rho^{2} -d^{2}u^{4}}} \, \partial_z - \rho \sqrt{\dfrac{1-u^{2}}{\rho^{2} - d^{2} u^{4}}} \, \partial_\rho \\
    \partial_\varphi
\end{pmatrix} .
\end{equation}

Applying the rotation defined in Eq.~\eqref{eq:3.35} to the given tetrad, we find a highly nonlinear partial differential equation for $L$, equivalent to the condition~\eqref{eq:2.20}, from which a new vacuum field can be obtained. Remarkably, we were able to solve the equation exactly, and found that the new tetrads from the solutions have exact $H^{\flat} + \Tilde{H}^{\flat}$. These solutions are
\begin{align}
    &L_{1} =  \frac{u \rho}{(1+\sqrt{1-u^{2}})\sqrt{\rho^{2} - d^{2} u^{2}}} \, , \label{eq:3.39}\\[1em]
    &L_{2} = -\frac{u}{2 \rho} \frac{\rho^{2} - d^{2}u^{2}(2-u^{2})}{\sqrt{(1-u^{2})(\rho^{2} - d^{2} u^{2})}} \,. \label{eq:3.40}
\end{align}

For each field defined by the new tetrads resulting from these transformations, we write the magnetic flux function as $\psi(v) = \psi_{0} v$. Then, the corresponding level set functions for the two new fields are
\begin{align}
    v_{1} &=  \frac{u \sqrt{\rho^{2} -d^{2}u^{2}}}{1+\sqrt{1-u^{2}}} \, , \label{eq:3.41}\\[1em]
    v_{2} &=  \frac{u^{3} \sqrt{\rho^{2} - d^{2} u^{2}}}{\rho^{2} - d^{2} u^{4}} \,. \label{eq:3.42}
\end{align}
One can easily verify that these new solutions indeed satisfy the vacuum stream equation in flat spacetime. As previously noted, these solutions were not found in Ref.~\cite{Compere_2016} using the foliation-dependent method, as the function builder the authors used cannot accommodate the functions \eqref{eq:3.41} and \eqref{eq:3.42}.

\begin{figure*}[ht]
\subfloat[asymptotically parabolic\label{fig:1.1a}]
{\includegraphics[width=0.45\linewidth]{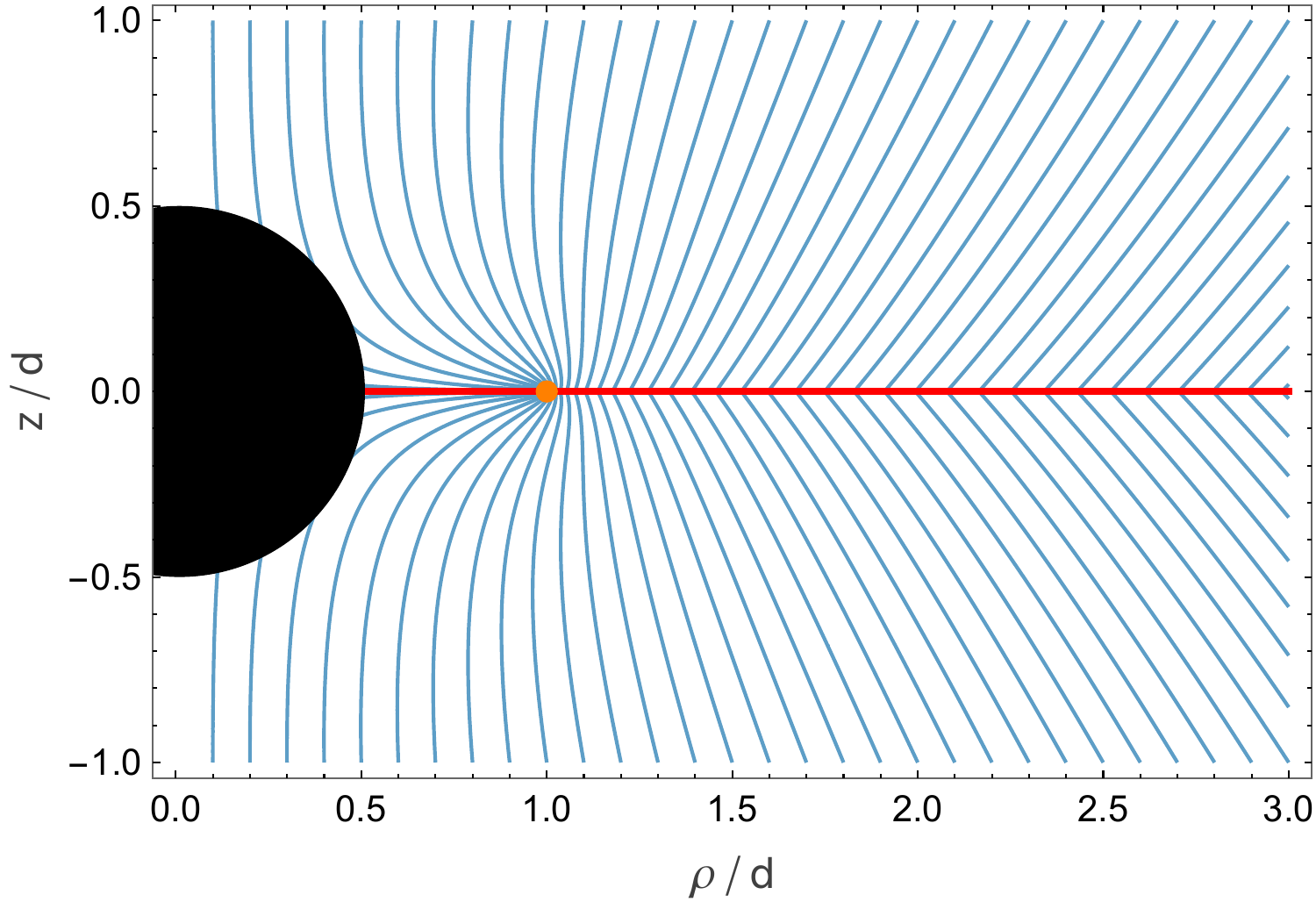}
}
\hfill
\subfloat[asymptotically dipolar\label{fig:1.1b}]
{\includegraphics[width=0.45\linewidth]{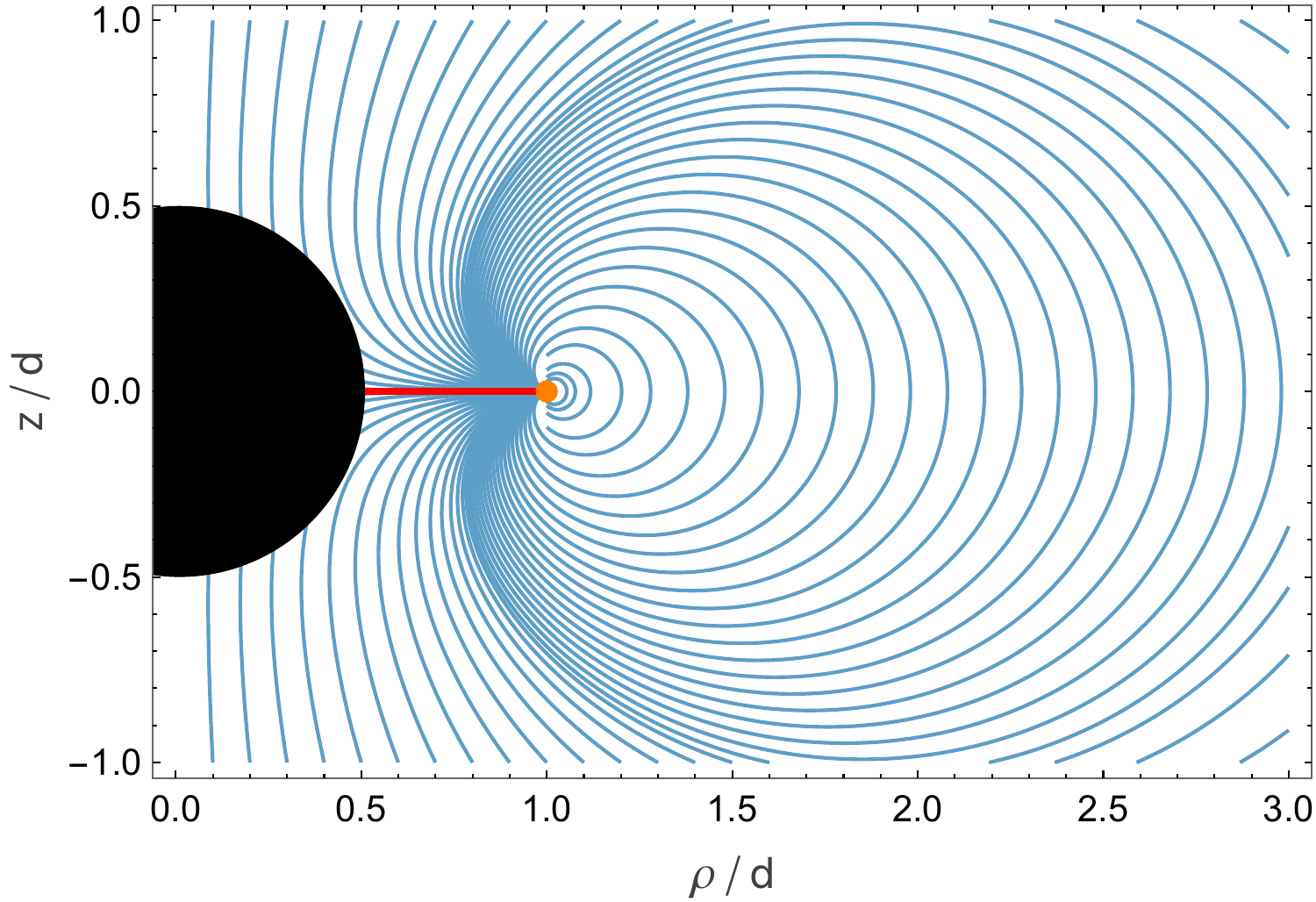}}
\caption{Poloidal magnetic field lines corresponding to (a) asymptotically parabolic and (b) asymptotically dipolar fields around a compact object (black-filled region). The solid red lines represent the thin disks, and the small orange circles mark the location of current concentrations.}
\label{fig:1.1}
\end{figure*}

The two flux functions are well-defined throughout flat spacetime and vanishes at the axis of symmetry, thereby preventing any line current there. It is obvious from $v_{1}$; we can rewrite and simplify the function $v_{2}$ further to obtain the expression that is also well-defined on the $z$-axis. However, unlike some known vacuum solutions in flat spacetime, we have not yet found exact rotating counterparts of these two solutions that describe pulsar magnetospheres. Nevertheless, the focus of this paper is on a black hole force-free field. 

As $d \rightarrow \infty$, the magnetic field lines of the external quadrupole field, or the X-point field\footnote{This field describes a magnetic null point.}, arises from the two new solutions,
\begin{equation}\label{eq:3.43}
    \lim_{d \rightarrow \infty} 2d^{2} v_{1} = \lim_{d \rightarrow \infty} d^{4} v_{2} = \rho^{2}z = r^{3} \sin^{2}{\theta} \cos{\theta} \,.
\end{equation}
On the other hand, the two solutions result in different magnetic field lines as $d \rightarrow 0$. In this limit, $v_{1}$ becomes parabolic, while $v_{2}$ becomes dipolar as follows:
\begin{equation}\label{eq:3.44}
\lim_{d \rightarrow 0} v_{1} = \sqrt{z^{2} + \rho^{2}} - z  \, ,\:\:\:
\lim_{d \rightarrow 0}
v_{2} = \frac{\rho^{2}}{(z^{2} + \rho^{2})^{3/2}} \,.
\end{equation}
These behaviors are visually evident in Fig.~\ref{fig:1.1}. From these plots, the feature of the limiting field line as $d \rightarrow \infty$, the external quadrupole field, is most apparent near the axis of symmetry whereas the structure of the limiting field line as $d \rightarrow 0$ becomes more prominent far from the axis. These happen because, in $\rho \ll d$ region, $d$ can be treated as large compared to the local distance, whereas $d$ can be treated as very small in the region $\rho \gg d$. For the same reason, the two new fields exhibit asymptotically parabolic and dipolar behavior at infinity, respectively. Hence, we refer to our new fields as the \textit{asymptotically parabolic} and \textit{asymptotically dipolar} fields in the remainder of the discussion.

In addition, in the limit $d \rightarrow \infty$, the two transformations given in Eqs.~\eqref{eq:3.39} and~\eqref{eq:3.40} correctly reduce to the appropriate transformation relating the tetrad of the vertical field, the asymptotic behavior of the hyperbolic field, to that of X-point field,
\begin{equation}\label{eq:3.45}
    \lim_{d \rightarrow \infty } L_{1} = \lim_{d \rightarrow \infty} L_{2} = - \frac{\rho}{2z} \, .
\end{equation}
Likewise, as $d \rightarrow 0$, the two transformations correctly reduce to the transformation from the radial field, which corresponds to the limiting field of the hyperbolic field, to the respective limits of the new solutions: from radial to parabolic,
\begin{equation}\label{eq:3.46}
    \lim_{d \rightarrow 0} L_{1} = \frac{\sqrt{z^{2} + \rho^{2}}-z}{\rho} \, ,
\end{equation}
and from radial to dipolar, 
\begin{equation}\label{eq:3.47}
    \lim_{d \rightarrow 0} L_{2} = -\frac{\rho}{2z} \,.
\end{equation}
These transformations are reported in Appendix~\ref{apx:A}.

\subsection{Source of the vacuum fields}\label{sec:IIIB}

Our new fields are sourced by disk current sheets at the equatorial plane, and both exhibit an apparent current singularity at $\rho = d$. In both cases, the pullback of $F$ to the plane is continuous, which implies the absence of magnetic monopoles. For the asymptotically parabolic solution, the surface current is given by 
\begin{align}
    &J = [\mathcal{J}_{1} H(\rho - d) + \mathcal{J}_{2}  H(d - \rho)]  \delta(z)  d z \wedge d\rho \wedge dt , \notag \\[1em]
    &\mathcal{J}_{1} :=  -\frac{ (\psi_{0}/\pi)}{\rho} \, , \:\: \mathcal{J}_{2} := \frac{ (\psi_{0}/\pi)\rho}{ d^{2} - \rho^{2} + d \sqrt{d^{2} - \rho^{2}}} \,,
    \label{eq:3.48}    
\end{align}
where $\delta(x)$ is the Dirac delta function and $H(x)$ is the Heaviside function (equals to one for $x>0$ and vanishes for $x<0$). It implies two discontinuous surface currents in the disk that behave differently: an inner current density, 
\begin{equation}\label{eq:3.49}
    \Vec{J}_{\text{inner}} = -\frac{( \psi_{0}/\pi)   \rho}{d^{2} - \rho^{2} + d \sqrt{d^{2} - \rho^{2}}} \delta(z) H(d - \rho) \, \hat{\varphi} \,,
\end{equation}
and an oppositely flowing outer current, 
\begin{equation}\label{eq:3.50}
    \Vec{J}_{\text{outer}}
    = \frac{ (\psi_{0}/\pi)}{\rho} \delta(z) H(\rho - b) \, \hat{\varphi} \, ,
\end{equation}
that are separated by a diverging current density at $\rho = d$ from the inner current. The divergence exhibits only an inverse square-root dependence on the separation distance, which is milder than the inverse-distance singularity characteristic of a line current. We can find then that the total current integrated over the disk is finite. The same behavior applies to the magnetic field. Thus, while the disk contains a localized concentration of current at $\rho = d$, this concentration is weaker than that associated with a line current.

On the other hand, in the asymptotically dipolar solution, there is no outer disk extending beyond $\rho = d$, but a diverging current density is also present at the `outer' disk edge $\rho = d$. The divergence grows inversely with $\sim (d^{2} - \rho^{2})^{3/2}$, faster than the divergence associated with a line current. It is unsurprising because $\rho = d$ is a caustic of magnetic field lines, which is visible in Fig.~\ref{fig:1.1b}. Consequently, the total current diverges, rendering this solution nonphysical. 

In both cases, the parameter $d$ does not represent an inner disk radius, as in the hyperbolic solution. One may redefine the solutions as $\psi^{\prime}=\sgn(z) \psi$, thereby restructuring the disks so that $d$ can be interpreted as the inner disk radius. However, this modification introduces magnetic monopoles on the disks, which is undesirable.

Moreover, in both solutions, the magnetic fields that cross a putative compact object, surrounded by the disk, intersect its surface twice (except for the last magnetic field touching the object). The double crossing of field lines on the surface and the divergence of the current density can be avoided in both cases by remodeling each solution. In each case, this is done by splitting the solution into northern and southern hemisphere parts and translating these local solutions along the $z$-axis,
\begin{equation}\label{eq:3.51}
    v(\rho,z) \rightarrow v^{\prime}(\rho,z) = v(\rho ,z+ \sgn(z) z^{\prime})  ,
\end{equation}
by an amount $z^{\prime}$ greater than the radius of the object, and gluing them together at $z = 0$. The solutions are still force-free because the flat spacetime stream equation admits translational symmetry along $z$-axis in general. However, this construction may not correspond to an exact solution in Kerr spacetime and is somewhat artificial, even if the perturbative solution is still possible, since the stream equation in Kerr spacetime lacks such translational symmetry. If one nevertheless insists that such a solution exists in Kerr spacetime, the field is sourced by a continuous disk that extends from the horizon and vanishes at infinity. Thus, in this paper, we exclusively consider the unmodified solution shown in Fig.~\ref{fig:1.1}, which exhibits a concentration of current on the disk.

\section{Black hole force-free jet}\label{sec:IV}

In the black hole magnetosphere, magnetic fields are only effectively tied to the horizon, or rather the
so-called stretched horizon, when viewed through the membrane paradigm~\cite{Thorne_Price_Macdonald_1986,Thorne:1982}. This horizon membrane acts like a unipolar inductor with finite resistance. For this reason, if the black hole is treated as the central object in our new force-free field, field lines that would otherwise cross the surface—here, the horizon—twice become segmented into two causally disconnected parts. The open field lines that extend to infinity constitute the black hole jet, while the closed field lines that connect to the magnetized disk allow for the energy exchange between the black hole and the disk. Such black hole-disk field lines are featured in numerical solutions~\cite{Uzdensky:2004,Mahlmann:2018}, and their possible physical realization has been proposed in the magnetosphere of a rotating black hole surrounded by a magnetized accretion torus~\cite{vanPutten:2002}. Both types of field lines must obey the regularity condition at the horizon, but are subject to different conditions at their other ends, and therefore rotate differently. However, in this paper, we focus exclusively on the properties of the black hole jet. The jet solution is obtained by promoting the asymptotically parabolic solution in flat spacetime to Schwarzschild background and performing a BZ perturbation, as discussed next.

Blandford and Znajek~\cite{BlandfordZnajek1977} introduced a perturbative method for solving the stream equation in the Kerr spacetime as a power series expansion in the black hole spin. The method begins by considering a seed vacuum degenerate field in Schwarzschild spacetime, with a given flux function $\psi = X(r,\theta)$. In this paper, our seed solution is obtained by promoting our new asymptotically parabolic vacuum field in flat spacetime, with the flux function
\begin{align}\label{eq:4.52}
     \psi_{\text{flat}} &=\psi_{0} \dfrac{u\sqrt{\rho^{2} -d^{2}u^{2}}}{1+\sqrt{1-u^{2}}}, \notag \\
     u &= \frac{1}{2d} \biggl[\sqrt{(\rho +d)^{2} + z^{2}} - \sqrt{(\rho -d)^{2} + z^{2}} \biggr],   
\end{align}
to the Schwarzschild spacetime, which will be discussed in Sec.~\ref{sec:IVA}. One then constructs a field with the seed flux function, and angular velocity and current given by
\begin{align}
    \Omega_{F} &= \dfrac{a}{M^{2}} W(X)  \, , \label{eq:4.53}\\[1em]
    I &= \dfrac{a}{M^{2}}Y(X)   \, ,  \label{eq:4.54}
\end{align}
where $Y, W$ are independent of the black hole spin. This field, with any functions of $Y$ and $W$, is force-free to first order in the black hole spin, $\mathcal{O}(a)$. But, this leaves us infinite degrees of freedom in $Y$ and $W$.

To obtain the unique solution, two appropriate equations relating the three functions are needed to fix $Y$ and $W$. These equations are provided by the regularity condition on the future horizon of the Kerr black hole $r_{H}$, and, in jet cases, by asymptotic condition at infinity. The functions $Y$ and $W$ are then fixed by imposing $I_{r_{H}} = I_{\infty}$ throughout the entire jet, since the total current enclosed by a poloidal loop is always conserved along the field lines in a force-free environment.

The first equation is provided by the Znajek horizon condition~\cite{Znajek:1977},
\begin{equation}\label{eq:4.55}
    I = (\Omega_{F}-\Omega_{H}) \, \sin{\theta} \, \partial_{\theta} \psi \:\:\:\:\: \text{at } r=r_{H} \, ,
\end{equation}
which is a universal relation that always holds on the black hole future horizon. Here, $\Omega_{H}$ is the angular velocity of the black hole horizon, which can be approximated as $\Omega_{H} \simeq a/(4M^{2})$. This condition arises from the regularity requirement of the field at the horizon.

On the other hand, there is no universal relation that can be imposed at infinity that, together with the Znajek condition, uniquely fixes the two unknown functions for a general field structure\footnote{Regularity of the jet solution at the light surfaces, where the stream equation becomes singular and particles co-rotating with the magnetic field lines flow at the speed of light, should be imposed. But the necessity of imposing this condition arises only upon consideration of higher orders in the perturbative expansion (see, for example, Refs.~\cite{Armas_2020,Camilloni:2022}).}. Hence, one is still left with an extremely large class of solutions, which may not correspond to a family of exact solutions in Kerr spacetime, as pointed out in Ref.~\cite{Gralla_etal_2016}. In the original BZ approach, this issue is addressed by proposing a postulate concerning the relationship among the three functions at infinity. It assumes that the asymptotic behavior of the perturbative solution at large $r$ follows the behavior of the rotating solution of the flux function in flat spacetime, since Kerr spacetime becomes approximately flat in this limit. This postulate, however, requires $\psi$ to admit a rotating counterpart in flat spacetime, even though the current and angular velocity in the perturbation are constructed as spin-induced quantities. Fortunately, there are mathematically consistent procedures for determining the appropriate conditions to impose at infinity for different field structures, without requiring exact rotating counterparts in flat spacetime under the perturbation scheme. The appropriate conditions can be derived by imposing regularity at infinity, equivalent to the Znajek condition~\cite{Nathanail_2014,Pan_2016}. These conditions, however, vary for different field structures, and apparent inconsistencies may arise for some fields when higher-order BZ perturbation is applied together with these conditions, which can nevertheless be resolved by other means.

For instance, Grignani \textit{et al.}~\cite{Grignani_2020} studied different types of asymptotics when the field is very close to the rotation axis in Kerr spacetime, where the force-free approximation is most reliable, under the standard BZ perturbation. In particular, they considered monopole-type, parabolic-type, and vertical-type asymptotics. bThey showed that asymptotically monopolar solutions contain higher-order expansions in spin that are divergent at infinity under the perturbation when the appropriate asymptotic condition corresponding to the field structure is imposed, agreeing with the previous reports~\cite{Grignani:2018,Tanabe:2008}. But, this issue can be resolved and a consistent higher-order solution can be achieved by allowing logarithmic contributions in $a$ in the perturbation and through the method of matched asymptotic expansions~\cite{Armas_2020,Camilloni:2022}. So, the asymptotically monopolar first-order jet solution of Ref.~\cite{Gralla_etal_2016} requires an appropriate matched asymptotic expansion scheme to avoid inconsistencies when extending the solution to higher orders in spin. This issue, however, does not arise for asymptotically parabolic fields under the perturbation, provided that the condition,
\begin{equation}\label{eq:4.56}
    I = \pm 2 \Omega_{F} \psi \:\:\:\: \text{at } r \rightarrow \infty  \, ,
\end{equation}
is met in first order. But, this equation is also the analogous Znajek condition for asymptotically parabolic fields at infinity~\cite{Pan_2016}. On the other hand, higher-order BZ expansion will always be valid for any asymptotically vertical fields, without requiring any further condition; the angular velocity and current are fixed by their asymptotic condition, identical to Eq.~\eqref{eq:4.56}~\cite{Pan_2016, Nathanail_2014}. Although we only consider a first-order solution in this paper, it is important to discuss these issues for the convenience of possible future work extending our solution to higher-order BZ perturbation, and to provide a strong justification for imposing the asymptotic condition used here.

In Sec.~\ref{sec:III}, we found two vacuum solutions in flat spacetime that may be promoted to Schwarzschild spacetime and may then serve as seed solutions for BZ perturbation. But only the asymptotically parabolic solution is viable as a physically valid solution in the Kerr spacetime. As mentioned in Sec.~\ref{sec:III}, the asymptotically dipolar solution has a caustic of magnetic field lines in flat spacetime located at the outer edge of the disk, which cannot be removed in the Kerr background. Because of this feature, the total current throughout the disk diverges and all the magnetic field lines corotate with the outer edge of the disk, and thus the Znajek condition cannot also be satisfied by the magnetic field threading the horizon in the black hole case. Moreover, a rotating black hole cannot sustain a magnetosphere consisting entirely of closed magnetic field lines sourced by an accretion disk, due to the strong twisting of the field lines, and the jet formation naturally involves collimation of plasma flowing along open field lines~\cite{Uzdensky:2004,Parfrey:2014, Bransgrove:2021}. Hence, the poloidal structure of a dipolar field in a nonrotating background cannot, in principle, be sustained around a rotating black hole, and thus the BZ perturbation cannot be performed consistently. For these reasons, we consider only the asymptotically parabolic solution to model a black hole jet in this paper and investigate its properties. The asymptotically dipolar solution may instead be relevant for magnetospheres of other astrophysical objects. Also, in what follows, our analysis is confined to the northern hemisphere, $0 \leq \theta \leq \pi/2 $, since the field in the southern hemisphere can be recovered by reflection.

\subsection{Magnetic flux function}\label{sec:IVA}

To promote the asymptotically parabolic solution in flat background to Schwarzschild spacetime, we first perform mode expansions of Eq.~\eqref{eq:4.52} around $r = 0$ and $r= \infty$, which we refer the corresponding expansions as the inner and outer solutions, respectively. The resulting expansions are the following:
\begin{align} 
\psi^{<}_{\text{flat}}(r,&\theta) = \frac{r}{\sqrt{\pi}} \sum^{\infty}_{k = 1} \frac{\Gamma(k+\frac{1}{2})}{\Gamma(k+1)} \biggl( \frac{r}{d} \biggr)^{2k} \, \Theta_{2k}(\theta) \, , \label{eq:4.57} \\[1em]
\psi^{>}_{\text{flat}}(r,& \theta) = r(1- \cos{\theta}) \notag \\
    & - \frac{d}{2\sqrt{\pi}} \sum^{\infty}_{k = 1} \frac{\Gamma(k-\frac{1}{2})}{ \Gamma(k+1)}  \biggl( \frac{d}{r} \biggr)^{2k-1} \Theta_{2k-1}(\theta) \,, \label{eq:4.58}
\end{align}
where we set $\psi_{0} =1$ in the meantime. Both series in the two expansions have a common radius of converge $d$, the location of the current concentration. Hence, $\psi^{<}_{\text{flat}}$ is valid only in $r \leq d$, while $\psi^{>}_{\text{flat}}$ is valid in $r \geq d$. It can be verified that these solutions agree on $r = d$, where they are equal to $d(\sqrt{\cos{\theta}}-\cos{\theta})$. Also, in these expansions, the asymptotically parabolic behavior of the solution becomes evident.

As discussed in Sec.~\ref{sec:II}, each mode in the expansions can be promoted to Schwarzschild spacetime by sending $r^{2k 
+ 1} \rightarrow R^{<}_{2k}(r)$ and $r^{-(2k-1)} \rightarrow R^{>}_{2k-1}(r)$ in the summations of Eqs.~\eqref{eq:4.57} and~\eqref{eq:4.58}, respectively. In doing so, the radius of convergence of both summations changes to
\begin{equation}\label{eq:4.59}
    d_{\circ} = d+M+\frac{M^{2}}{4d}\, ,
\end{equation}
which is now the location of the sign-reversal of the current disk and current concentration in the Schwarzschild solution~(see Appendix~\ref{apx.B}). Also, continuity of the promoted expansions at the new radius of convergence is ensured by introducing a relative normalization $C$ for the inner solution, determined in Appendix~\ref{apx.B}.

The flux function vanishes at the location of the current concentration in the disk in flat spacetime. Upon promoting the inner solution to Schwarzschild spacetime, the flux function continues to vanish at the location of the current concentration, as expected. However, the leading term in the outer solution~\eqref{eq:4.58} becomes $r+2M \log{(1- r/2M)}$ if one attempts the mapping $r \rightarrow R^{<}_{0}(r)$, which is a solution of the $l = 0$ radial equation. With this term, the promoted outer expansion does not vanish at the location of the current concentration. To address this issue, we determined the general solution of the $l = 0$ radial equation. The constants of integration are then fixed by requiring the full outer flux function to vanish at the location of the current concentration, and by requiring the solution to reduce to $r$ when $M = 0$, or as $r \rightarrow \infty$. The $\log{(r-2M)}$ term in the solution does not pose a problem for the outer solution, provided that the location of the current concentration lies away from the horizon. The explicit expressions for the promoted inner and outer solutions are given in Eq.~\eqref{eq:apxB}.

\subsection{Current and angular velocity}

To implement the BZ perturbation, only the near-horizon behavior of the solution is required in our case. From Eq.~\eqref{eq:apxB.1a}, this behavior is
\begin{equation}
X(r=r_{H},\theta) \approx \alpha X_{H} \cos{\theta} \sin^{2}{\theta} , \label{eq:4.60}
\end{equation}
where
\begin{equation}\label{eq:4.61}
X_{H} =\frac{4  \,C X_{0} M^{3}}{\alpha\left[d_{\circ}-M+\sqrt{d_{\circ}(d_{\circ}-2M)}\right]^{2}}
\end{equation}
is the total magnetic flux of the jet on the horizon, balanced by a negative flux associated with black hole-disk field lines that results in a vanishing net magnetic flux on one hemisphere of the horizon. Here, $X_{0}$ is a constant representing the total magnetic flux per unit radial distance in one hemisphere at large $r$, and
\begin{equation}
\alpha = \csc^{2}{\theta_{\circ}} \sec{\theta_{\circ}} \label{eq:4.62}
\end{equation}
is a constant determined by the angle $\theta_{\circ}$ at which the last field line touches the horizon. This angle can be determined from the condition that $\partial_{\theta}\psi = 0$ at the horizon.

In the flat spacetime solution, the last field line touches a central object with radius $R$ at the angle
\begin{equation}
    \theta^{\text{flat}}_{\circ} = \arctan{\sqrt{\dfrac{2l^{6} + 2 l^{4} - 6l^{2} + 1+ \sqrt{9-8l^{2}}}{2(1-l^{2})^{3}}}} \, , \label{eq:4.63}
\end{equation}
where we define $l \equiv R/d$ and write the radicand in a form that explicitly requires $0 \leq l < 1$. In an external quadrupole field, the angle at which the last field line touches the surface is exactly $\arctan(\sqrt{2})$, or about $54.7^{\circ}$. For our solution, in the large-$d$ limit, the angle is
\begin{equation}\label{eq:4.64}
    \theta^{\text{flat}}_{\circ} = \arctan(\sqrt{2})+ \mathcal{O}(l^{2}) \, .
\end{equation}
The correction appears at second order in $l$ and higher. We expect that, in Schwarzschild background, the correction due to the black hole appears at the same order in $r_{H}/d_{\circ}$ in the limit $d_{\circ} \gg r_{H}$. Thus, we may approximate $\theta_{\circ} \simeq \arctan{(\sqrt{2})}$ in Eq.~\eqref{eq:4.62} in the large-$d_{\circ}$ limit.

However, the location at which the reversal of the disk current and the current concentration is most likely to happen is not far from the horizon, due to possible interesting dynamics between the strong gravitational and electromagnetic forces acting on the plasma in this region. Such an interaction may also be the reason a current disk persist inside the innermost stable circular orbit (ISCO). This behavior is observed in several analyses, such as simulations of magnetically arrested disks (MADs), where the inflow of the magnetized accreting matter is halted because of the concentration of magnetic field, and magnetic reconnection~\cite{Ripperda_2020, Igumenshchev_2008}. Hence, although results in the large-$d_{\circ}$ limit are analytically tractable and reliable, it is also necessary to analyze the small-$d_{\circ}$ regime. 

But the absence of a closed-form expression for the promoted magnetic flux expansions prevents us from deriving a general analytic expression for the angle $\theta_{\circ}$. We therefore rely exclusively on numerical computations to determine $\theta_{\circ}$. In Appendix~\ref{apx.C}, we describe the numerical scheme used to obtain the precise value of $\theta_{\circ}$ and show that $\theta_{\circ} = \arctan(\sqrt{2})$ in the limit $d_{\circ} \rightarrow \infty$, in agreement with our previous assumption. Moreover, the numerical computation shows that $\theta_{\circ}$ is close to $\theta_{\circ}^{\text{flat}}$ when $d_{\circ}$ is far from the horizon. The physical intuition behind this is that, since the last jet field line is also tied to the disk, the flux value of the last field line remains close to its flat-spacetime counterpart when $d_{\circ}$ is not close to the horizon. A large deviation would otherwise imply that the black hole significantly alters the structure of the last field line in a region where gravitational effects should be small. For instance, when $d_{\circ}$ is at the ISCO radius, $r = 6M$, the numerical approximation gives $\theta_{\circ} \approx 54.9^{\circ}$, while the flat-spacetime value using Eq.~\eqref{eq:4.63} is $\theta_{\circ}^{\text{flat}} \approx 56.5^{\circ}$. However, this approximation breaks down when $d_{\circ}$ approaches the horizon, where strong gravitational effects become significant along the entire last jet field line.

We can use Eq.~\eqref{eq:4.60} to treat the angle at the horizon $\theta_{H}$ as a parameter labeling the field lines,
\begin{equation}\label{eq:4.65}
    \theta_{H}(x) = \arccos{\biggl\{\frac{2}{\sqrt{3}} \cos{ \biggl[ \frac{1}{3} \arccos{\biggl(- \frac{3\sqrt{3}x}{2 \alpha} \biggr) } \biggr]} \biggr \}} ,
\end{equation}
where $x \equiv X/X_{H}$. Then, the Znajek horizon condition requires
\begin{equation}\label{eq:4.66}
    Y = \frac{\alpha}{2} X_{H} \biggl(W - \frac{1}{4} \biggr) [1+3 \cos{(2\theta_{H})}] \sin^{2}{(\theta_{H})} .
\end{equation}
Meanwhile, we use the condition in Eq.~\eqref{eq:4.56} at infinity,
\begin{equation}
    Y = -2WX , \label{eq:4.67}
\end{equation}
with the choice of sign as the appropriate sign to describe energy extraction by the jet. By equating Eqs.~\eqref{eq:4.66} and~\eqref{eq:4.67}, we find the angular velocity as
\begin{equation}
    W = \frac{1}{4} - \frac{x}{4x+\alpha [1+3  \cos{(2\theta_{H})}]\sin^{2}{\theta_{H}}} . \label{eq:4.68}
\end{equation}

\begin{figure}[ht]
    \centering
    \includegraphics[width=0.97\columnwidth]{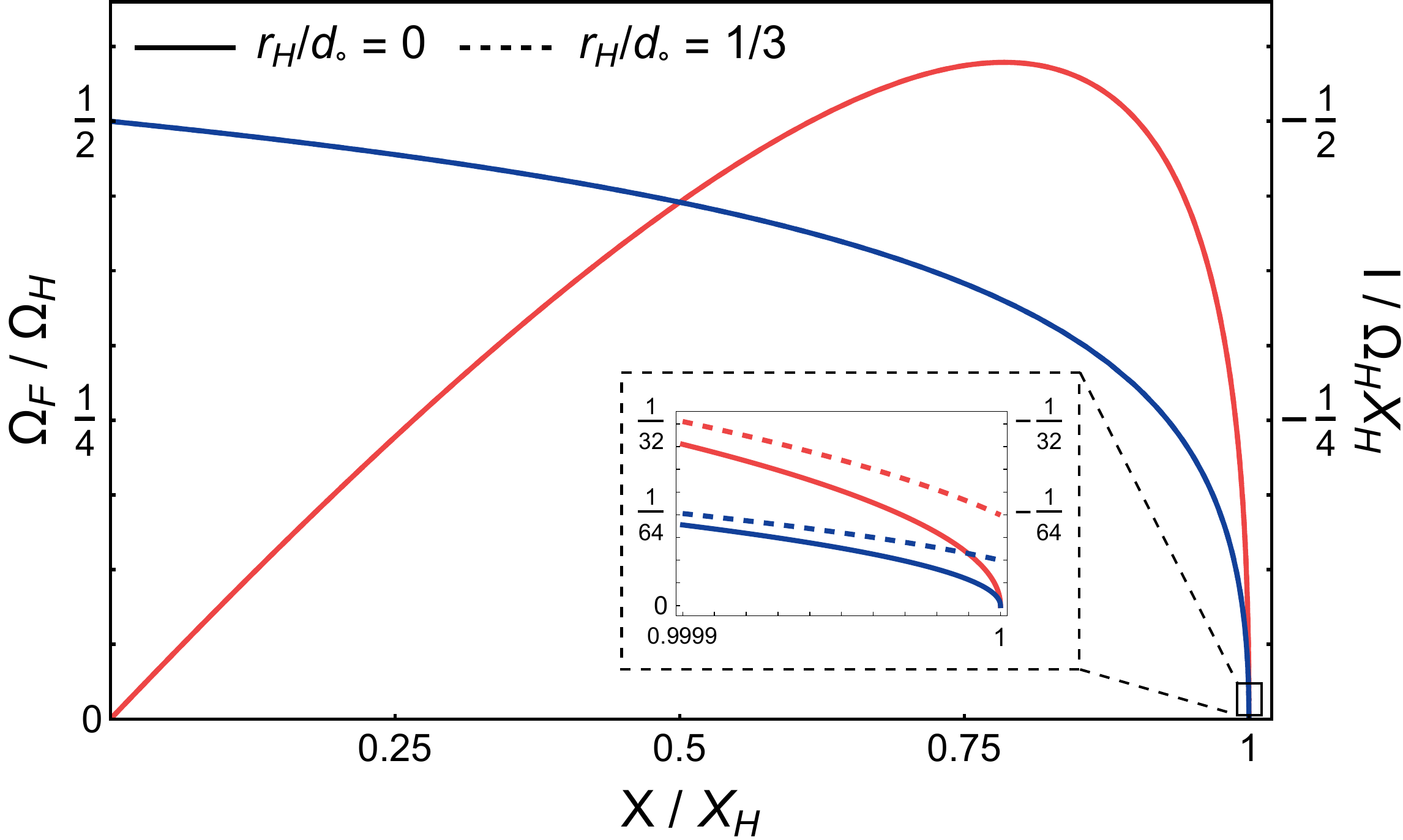}
    \caption{Angular velocity (blue) and current (red) of the field lines for $d_{\circ} = r_{\text{ISCO}}$ (dashed lines) and $d_{\circ} \rightarrow \infty$ (solid lines) as functions of $X/X_{H}$. The angular velocity and current are insensitive to $d_{\circ}$.}
    \label{fig:2}
\end{figure}

Judging from Eq.~\eqref{eq:4.68}, the angular velocity of the field lines is maximal along the rotation axis, where it is equal to half of the horizon angular velocity, and decreases monotonically toward the last field line. This angular velocity profile is evident in the cases we consider, $d_{\circ} = r_{\text{ISCO}} = 6M$ and $d_{\circ} \rightarrow \infty$, as shown in Fig.~\ref{fig:2}. We may also infer from the figure that, within this $d_{\circ}$ domain, the current and angular velocity are largely insensitive to $d_{\circ}$ over the bulk of the jet, with minimal differences appearing only near the boundary of the jet, at fixed $X_{H}$. The reason of this insensitivity is because $\theta_{\circ}(d_{\circ}\rightarrow \infty) \approx \theta_{\circ}(d_{\circ}= r_{\text{ISCO}})$ leading to $\alpha(d_{\circ}\rightarrow \infty) \approx \alpha( d_{\circ} = r_{\text{ISCO}} ) \approx 2.6$. We checked that this insensitivity also persists even when $r_{H} \ll d_{\circ}< r_{\text{ISCO}}$, where our numerical approximation is still reliable.

A difference arises for the last jet field line. The angular velocity of the last field line vanishes as $d_{\circ} \rightarrow \infty$, whereas it takes a nonzero value for $d_{\circ} = r_{\text{ISCO}}$; a similar behavior is observed in the current. This difference is largely a consequence of our approximation. In fact,
the last field line is expected to have zero angular velocity and current for any value of $d_{\circ}$. In the last field line, $\partial_{\theta}\psi$ vanishes at the horizon, which, together with the horizon regularity and asymptotic conditions, implies zero angular velocity and current. Also, since the last field line is also tied to a current in the disk, it should not corotate with the disk. Instead, the field line must slip through the disk, implying that the disk must have a finite resistance to account for the vanishing of its angular velocity.

In addition, the last jet field line has adjacent field lines: one anchored in the disk and extending to infinity, and another anchored in the disk and connected to the black hole. The field line anchored in the disk and extending to infinity may constitute the disk wind. Since the current and angular velocity of the last field line do vanish for any $d_{\circ}$, and since the same asymptotic condition as in the jet cannot be imposed on the disk-wind field line, a current sheet may possibly form between the jet and the wind. Meanwhile, there may exist, in principle, a transition region extending from the horizon to the disk current at which the last jet field line is slipping through, within which the angular velocity and current changes smoothly to zero. Moreover, the current along the black hole-disk field lines flows opposite to that of the jet field lines. This reversal happens because $\partial_{\theta}\psi$ of the hole-disk field lines changes sign, implying that energy flow from the disk to the black hole, as expected on physical grounds. But, whether the black hole-disk system as a whole results in net energy extraction depends on the properties of the hole-disk field lines and, ultimately, by the properties of the disk.
\vspace{-1em}
\subsection{Jet power and circuit model}

The Blandford–Znajek mechanism enables the extraction of energy from a rotating black hole by a force-free field, providing a unified explanation for the release of black hole electromagnetic jets. Specifically, a stationary axisymmetric force-free field around a rotating black hole admits Noether currents associated with the black hole isometries~\cite{BlandfordZnajek1977,Gralla_Jacobson:2014}. The current associated with the asymptotic time-translation symmetry corresponds to the \textit{Killing energy} (i.e., the energy measured at infinity), while the current associated with the azimuthal-translation symmetry corresponds to the \textit{Killing angular momentum}. The Killing energy and angular momentum are conserved as they propagate along the poloidal field lines. For a local observer, the Killing energy is interpreted as spatial angular momentum in the ergoregion. Hence, a negative energy flux can flow into the black hole from the ergoregion while being balanced by a positive energy flux escaping to infinity. The power output of the BZ process is given by
\begin{equation}\label{eq:4.69}
\mathcal{P} = - \frac{1}{2\pi} \int I(\psi) \, \Omega_{F}(\psi) \,  d\psi \, .
\end{equation}

We find that, in the large $d_{\circ}$-limit and for the $d_{\circ} = r_{\text{ISCO}}$, the power emitted by the asymptotically parabolic jet is practically the same,
\begin{equation}\label{eq:4.70}
 \mathcal{P} \approx 2.2 \times 10^{-2} X^{2}_{H} \Omega^{2}_{H}  \,.
\end{equation}
We note that this expression for the power accounts only for the jet in the northern hemisphere. This power is also comparable to the energy output of the hyperbolic jet at fixed horizon magnetic flux and angular velocity. The similarity of the power for the two disk parameter values, which is also observed in the hyperbolic jet~\cite{Gralla_etal_2016}, is attributed to the weak dependence of the angular velocity and the current on $d_{\circ}$. In general, however, the power differs significantly for different values of $d_{\circ}$ once the condition of fixed $X_{H}$ is relaxed, since $X_{H}$ depends on the disk. In particular, at finite $X_{0}$, the horizon magnetic flux becomes very small in far $d_{\circ}$, and the power correspondingly becomes insignificant too. This behavior is physically reasonable, as the disk that sources the magnetic field is then very far from the black hole. But, regardless of the disk profile, the power in~\eqref{eq:4.70} shows a clear interpretation of the BZ mechanism as an energy-extraction process on the horizon, in which the extracted energy depends solely on the flux of field lines threading the horizon and on the horizon angular velocity.

Also, another important aspect of our analysis at fixed $X_{H}$ is determining the effective resistance of the jet when it is modeled as a circuit with the load at infinity. This quantity is independent of the horizon magnetic flux and therefore provides a characterization of the jet that remains approximately valid for any value of $d_{\circ}$. The voltage drop across the field lines is given by the integral of $\Omega(\psi)\, d\psi$. In far $d_{\circ}$, the net current increases from zero to a maximum at $X/X_{H} \approx 0.785$, and then decreases to zero at the boundary of the jet. We may regard $X \lesssim 0.785\, X_{H}$ as the flow of forward current in the bulk of the jet and $X \gtrsim 0.785\, X_{H}$ as the return flow of current. The corresponding approximate bulk current of the jet is
\begin{equation}\label{eq:4.71}
   I(0.785 X_{H}) = I_{\text{bulk}} \approx -0.55\, X_{H}\Omega_{H},
\end{equation}
and the voltage drop across the jet is
\begin{align}\label{eq:4.72}
    V &= 4X_{H}\Omega_{H} \int^{1}_{0} W\!\left(x;\alpha = \frac{3 \sqrt{3}}{2} \right) dx \notag \\
      &\approx 0.405\, X_{H}\Omega_{H}. 
\end{align}
Then, effective resistance of the jet circuit is 
\begin{equation}
    R_{\text{eff}} = \frac{V}{|I_{\text{bulk}}|} \approx  0.74 \,. \label{eq:4.73}
\end{equation}
Since we find that the jet properties are practically independent on $d_{\circ}$, this resistance value holds for any $d_{\circ}$. Also, the effective resistance of our jet circuit is comparable to that of the black hole hyperbolic jet model, and both are greater than that of the stellar hyperbolic jet (see Ref.~\cite{Gralla_etal_2016}). Hence, our new analytic jet solution is consistent with the intuition of the membrane paradigm that black hole horizon provides additional effective resistance, as suggested by comparison with the stellar hyperbolic jet, although a stellar counterpart of our solution is not presented here.

\section{Summary and Outlook}

With the advances in our understanding of the geometric spacetime structure of force-free fields, we now have new tools and perspectives to search for analytic solutions to the force-free equations in curved spacetimes beyond conventional methods.

In this work, we combined these geometric innovations with previous methods to construct a new analytic jet model from a disk-fed rotating black hole. First, we found new classes of parametrized solutions in flat spacetime by applying the geometric procedure developed in Ref.~\cite{Adhikari_etal_2024} to a hyperbolic solution. Among these solutions, we identify one that is physically viable and of particular interest, while the other is deemed unphysical; we reported the `unphysical' solution anyway for its potential application to stellar magnetospheres.
Then, we extended the physical solution to Schwarzschild spacetime and seeded it into the Blandford-Znajek perturbative method, in an equivalent fashion to the construction of the black hole hyperbolic jet~\cite{Gralla_etal_2016,Beskin:2010iba}. Thus, our jet model constitutes a perturbative analytic solution.

Our black hole jet exhibits a radially varying structure, transitioning from a magnetic null field structure near the horizon to an asymptotically parabolic configuration at large distances. This variability is of interest, given that the observed jet boundary of $\mathrm{M87}^{*}$ displays approximately, rather than genuine, parabolic geometry at $\sim 10^{1}-10^{5}$ gravitational radius~\cite{Nakamura_2018}. It would be worthwhile to assess the jet boundary geometry of our model in future work. 

The system considered in our jet model corresponds to a black hole–disk system in which a thin equatorial disk, containing a coincident current sign reversal and current concentration, sources the magnetic field. The poloidal field lines extending from the horizon to infinity constitute the jet, while some other field lines thread the horizon and are anchored to the disk, allowing energy transport from the disk to the black hole. The remaining field lines beyond the jet remain anchored to the disk and may contribute to a magnetized disk wind, suggesting the possible formation of a current sheet at the jet boundary.

Moreover, the location of the current concentration and sign reversal parametrizes the flux function of our black hole jet, and so the total poloidal current and angular velocity. But, the jet properties are insensitive to this parameter at fixed horizon angular velocity and magnetic flux. Hence, we obtain a jet power given in Eq.~\eqref{eq:4.70} that is valid for any value of the parameter within the approximation. This situation is similar to that of the hyperbolic jet. In both the hyperbolic and our new asymptotically parabolic jets,
the insensitivity to the disk parameter extends not only to the jet power but also to the entire jet properties, although this aspect of the hyperbolic jet has not been strongly emphasized in previous literature~\cite{Gralla_etal_2016}. It would be interesting to investigate, however, whether the apparent insensitivity of the jets with the disk parameter reflects a genuine universality of the jets properties to the disk structures or an artifact of the approximations used in the slowly rotating black hole. Such degeneracy of the Blandford-Znajek jet power at small black hole spin has been observed in alternative theories of gravity, but it breaks at higher spin~\cite{Dong:2022,Camilloni:2024}.

Our results also illustrate the standard interpretation that the Blandford-Znajek mechanism operates as an energy-extraction process at the horizon, independent of the source of the magnetic field. By examining the jet power derived here, together with those of other jets from disk-fed slowly rotating black holes, one finds that all jet powers depend only on the magnetic flux threading the horizon and the horizon angular velocity, and not on the detailed properties of the disk --- a feature expected from the BZ process. The question of whether this process should be understood as being mediated by the horizon through the membrane paradigm or by the ergosphere has been extensively debated. But numerical simulations favor the latter interpretation~\cite{Ruiz_2012, Toma_2025}.

In addition, the effective resistance of the jet circuit, when compared with that of a known stellar jet model, is compatible with the membrane paradigm notion that the horizon possesses a finite resistance. It is also comparable to the effective resistance found in the black hole hyperbolic jet.

\begin{acknowledgments}
LV acknowledges the financial support
of the Philippines’ Department of Science and Technology
through the Advanced Science and Technology Human Resources Development Program.
\end{acknowledgments}

\appendix
\section{Transformation of vacuum tetrad}\label{apx:A}

In this section, we show that tetrads associated with vacuum degenerate fields in spherically symmetric spacetimes, 
\begin{equation}\label{eq:A1}
    ds^{2} = - g_{tt} dt^{2} + g_{rr} dr^{2} + r^{2} d\theta^{2} + r^{2}\sin^{2}{\theta} \, d\varphi^{2} ,
\end{equation}
such as flat spacetime, are related by orthogonal transformations.

In general, a vacuum degenerate field in such spacetime can always be written as 
\begin{equation}\label{eq:A2}
    F = h\, d\hat{\psi} \wedge \omega  \, , \:\:\: h \equiv \dfrac{\| d\psi \|}{r\sin{\theta}} ,
\end{equation}
where $d\hat{\psi} \equiv d\psi/ \| d\psi \|$ and $\omega \equiv (r\sin{\theta}d \varphi)$. The flux function depends only on the poloidal coordinates $(r,\theta)$, so $d\hat{\psi}$ lies on the poloidal subspace that is constant along the integral curves of the Killing fields $\partial_{t}$ and $\partial_{\varphi}$, and orthogonal to $\omega$. In contrast, $\omega$ lies on the toroidal subspace generated by the Killing fields with coordinates $(t, \varphi)$. 

Using the identity for the dual of a $p$-form provided in the Appendix 2.2 of Ref.~\cite{Gralla_Jacobson:2014}, the dual field is
\begin{equation}\label{eq:A3}
    *F = h \upsilon \wedge \rho \, ,
\end{equation}
where $v \equiv  -\sqrt{g_{tt}} dt$ and $\rho \equiv \star \,  d\hat{\psi} / \sqrt{g_{tt}}$, with $\star$ denoting the dual taken with respect to the poloidal subspace. These one-forms, $\upsilon$, $\rho$, $d\hat{\psi}$, $\omega$, make a tetrad in the background spacetime. Thus, we can assign them as $e^{\flat}_{0}$, $e^{\flat}_{1}$, $e^{\flat}_{2}$, $e^{\flat}_{3}$, respectively. Being a degenerate Maxwell field means that $e_{0}$ and $e_{1}$ span an involutive distribution. 

Notably, $v$ and $\omega$ depend only on the background geometry and not on the magnetic flux function. Then, $e^{\flat}_{0}$ and $e^{\flat}_{1}$ are always the same for a vacuum field on a fixed background spacetime. Hence, any tetrad associated with a vacuum field and transformed by a homogeneous Lorentz transformation satisfies
\begin{equation}\label{eq:A4}
    \Bar{e}_{0} =\Lambda^{a}_{\:\:0}e_{a} = e_{0} , \:\:\:\: \Bar{e}_{3} =\Lambda^{a}_{\:\:3}e_{a} = e_{3} \, ,
\end{equation}
which implies that 
\begin{equation}\label{eq:A5}
    \delta_{ij}\Lambda^{j}_{\:\: 0} = \delta_{i0} \, , \:\:\: 
    \delta_{ij}\Lambda^{j}_{\:\: 3} = \delta_{i3} \, ,
\end{equation}
where $\delta_{ij}$ is the Kronecker delta. Moreover, with this condition and the similarity transformation $\Lambda^{T} \eta \Lambda = \eta$, we also have the identities
\begin{equation}\label{eq:A6}
\delta^{ij}\Lambda^{0}_{\:\: j} = \delta^{i0}  , \: \delta^{ij}\Lambda^{3}_{\:\: j} = \delta^{i3}  , \:
\delta_{kl} \Lambda^{k}_{\:\: m} \Lambda^{l}_{m^{\prime}} = \delta_{m m^{\prime}} \,,
\end{equation}
with $k,l,m,m^{\prime} =1,2$. But this is exactly the definition of orthogonal transformation on $e_{1}$-$e_{2}$ plane, or on the poloidal plane.  For instance, the vertical, X-point, radial, parabolic and dipolar fields have the following respective tetrad bases:

\begin{tabular}{ll}
& \\
All fields: & \begin{tabular}{l}
        $e_{0} =  \partial_{t} \, $,     
         $e_{3} = \dfrac{1}{\rho}\partial_{\varphi}  \, $
    \end{tabular} ; \\
Vertical: & \begin{tabular}{l}
        $e_{1} = - \partial_{z}\, $,
         $e_{2} = \partial_{\rho} \, $
    \end{tabular} ; \\
X-point: & \begin{tabular}{l}
        $e_{1} = \dfrac{\rho }{\sqrt{4z^{2}+\rho^{2}}} \partial_{\rho}- \dfrac{2z}{\sqrt{4z^{2}+\rho^{2}}}\partial_{z}  $   \\
         $e_{2} = \dfrac{\rho}{\sqrt{4z^{2}+\rho^{2}}} \partial_{z} + \dfrac{2z}{\sqrt{4z^{2}+\rho^{2}}} \partial_{\rho}   $
    \end{tabular}  ; \\
Radial: & \begin{tabular}{l}
        $e_{1} = - \dfrac{z}{\sqrt{z^{2} +\rho^{2}}} dz - \dfrac{\rho}{\sqrt{z^{2}+ \rho^{2}}} \partial_{\rho}\, $   \\
         $e_{2} = - \dfrac{\rho}{\sqrt{z^{2} + \rho^{2}}} \partial_{z} + \dfrac{z}{\sqrt{z^{2}+\rho^{2}}} \partial_{\rho}\, $ 
    \end{tabular} ; \\
Parabolic: & \begin{tabular}{l}
        $e_{1} =\dfrac{\left[ (z-\sqrt{z^{2}+\rho^{2}})\,\partial_{\rho}- \rho \, \partial_{z} \right]}{\sqrt{2(z^{2} +\rho^{2}-z\sqrt{z^{2}+\rho^{2}}})} $   \\[1em]
         $e_{2} = \dfrac{\left[(z-\sqrt{z^{2}+\rho^{2}})\,\partial_{z} + \rho \, \partial_{\rho} \right]}{\sqrt{2(z^{2} +\rho^{2}-z\sqrt{z^{2}+\rho^{2}}})} $ 
    \end{tabular} ; \\
Dipolar: & \begin{tabular}{l}
        $e_{1} =\dfrac{\left[(\rho^{2}-2z^{2}) \, \partial_{z} - 3 z\rho \,\partial_{\rho}\right]}{\sqrt{(z^{2}+ \rho^{2})(4z^{2}+ \rho^{2})}}\, $   \\[1em]
         $e_{2} = \dfrac{\left[ - 3 z\rho \,\partial_{z}+(2z^{2}-\rho^{2}) \, \partial_{\rho}\right]}{\sqrt{(z^{2}+ \rho^{2})(4z^{2}+ \rho^{2})}}\, $ 
    \end{tabular} .
\end{tabular}
\vspace{0.25cm}

\noindent
With these tetrads, it can easily be shown that the vertical tetrad transforms into X-point tetrad when the transformation in Eq.~\eqref{eq:3.35} is applied with $L = \rho/2z$. With a similar transformation, the radial tetrad transforms into parabolic and dipolar tetrads for $L = (\sqrt{z^{2}+ \rho^{2}}-z)/\rho$ and $L = - \rho/2z$, respectively. We find that all vacuum solutions in Ref.~\cite{Compere_2016} are related by this transformation.
\section{Normalization and radius of convergence of Schwarzschild solution}\label{apx.B}

The promoted inner and outer flux functions in Schwarzschild spacetime referred to in Sec.~\ref{sec:IVA} are
\begin{subequations}\label{eq:apxB}
\begin{empheq}{align}
&\psi^{<}(r,\theta) = \frac{C}{\sqrt{\pi}} \sum^{\infty}_{k =1} \frac{\Gamma(k+\frac{1}{2})}{\Gamma(k+1)} \frac{R^{<}_{2k}(r)}{d^{2k}} \, \Theta_{2k}(\theta) \, , \label{eq:apxB.1a} \\
&\psi^{>}(r,\theta) = \biggl[ r - d_{\circ}+ 2M \log\biggl( \frac{r - 2M}{d_{\circ}- 2M} \biggr) \notag \\ 
&+ \frac{1}{2\sqrt{\pi}} \sum^{\infty}_{k = 1} \frac{\Gamma(k - \frac{1}{2})}{\Gamma(k+1)} d^{2k} R^{>}_{2k-1}(d_{\circ}) \biggl](1-\cos{\theta} ) \notag \\ 
&- \frac{1}{2\sqrt{\pi}} \sum^{\infty}_{k = 1} \frac{\Gamma(k - \frac{1}{2})}{\Gamma(k+1)} d^{2k}R^{>}_{2k-1}(r) \Theta_{2k -1}(\theta) \, , \label{eq:apxB.1b}    
\end{empheq}
\end{subequations}
where we may regard $d_{\circ} > 2M$ as the new free parameter.

As mentioned in Ref.~\cite{Gralla_etal_2016},
the series involved in any promoted solutions are expected to match because of the unique, well-defined correspondence between the solutions in the flat and Schwarzschild spacetimes. But, in the promotion we employed, the association acts only on local neighborhood of the exact global solution up to an undetermined overall constant. As a consequence, promoting any flat spacetime solution in this way leaves a relative normalization between the `local' promoted solutions. In addition, the matching of the inner and outer solutions at the same radius of convergence $d_{\circ}$, for all the angles can be assessed only numerically. In particular, we can check that the solutions match at $d_{\circ}$ using the expression for $C$ derived from the angle $\theta = \pi/4$ in this section. A definitive numerical verification is hindered, however, by the slow convergence of the series for any angle. In fact, the series falls only like $k^{-3/2}$ for $\theta = \pi /4$, and it
\begin{figure}[h!]
    \centering
    \includegraphics[width=0.95\columnwidth]{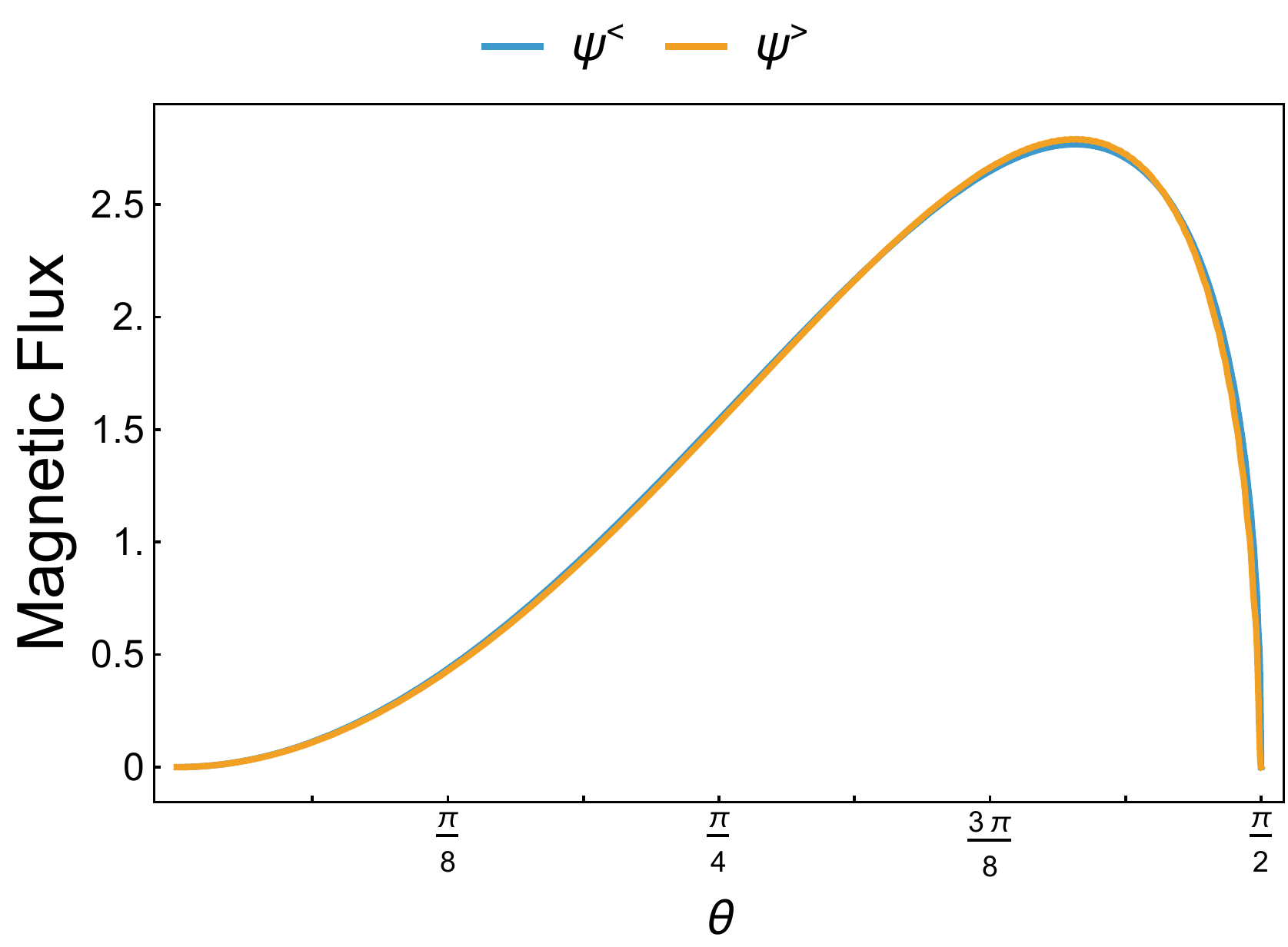}
    \caption{Inner (blue) and outer (orange) magnetic flux functions evaluated at $d_{\circ}$ for all the northern polar angles all using the same normalization $C$ derived in Appendix~\ref{apx.B}. We used $d_{\circ} = 6M$ and $M =2$ in this plot.}
    \label{fig:3}
\end{figure}
requires $\sim 10^{3}$ summation terms to obtain a good accuracy. 
We have shown the matching of the two solutions, anyway, at a specific condition in Fig.~\ref{fig:3} with reasonable precision. We have therefore relied on the presumed continuity of the solutions in any circumstances from this example, along with the qualitative verification of our expression of $C$ at the end of this section.

In this section, we show that the common radius of convergence of the solutions, implied by the present summations, is $d_{\circ}$, and we derive an approximate expression for $C$. We do not derive the radius of convergence for the flat spacetime solution here, as it is obtained by taking the $M = 0$ limit of the result derived in this section. We also expect it to be equal to $d$, since the solutions were, from the outset, expansions inside and outside the radius $d$. Although this result can be proved analytically in a straightforward manner, we omit the proof in this section. Throughout this section, we make use of the asymptotic approximations of the hypergeometric function given on p. $77$ of Ref.~\cite{bateman_1953_cnd32-h9x80},
\begin{widetext}
\begin{align}
    &\phantom{}_{2}F_{1}[a+ \lambda, a - c + 1+ \lambda; a- b + 1 + 2 \lambda; 2(1-z)^{-1}] \overset{\lambda \rightarrow \infty}{\sim}
    2^{a+b} 
    \left(\frac{z-1}{2}\right)^{a + \lambda} \dfrac{\Gamma(a-b+1+2\lambda)\Gamma(\frac{1}{2}) \lambda^{-1/2}}{\Gamma(a-c+1+ \lambda) \Gamma(c-b+ \lambda)} 
    \notag \\   
    &\times e^{-(a+ \lambda) \xi}  (1- e^{-\xi})^{-c+ \frac{1}{2}} (1+ e^{-\xi})^{c - a - b - \frac{1}{2}} [1+ \mathcal{O}(\lambda^{-1})] \, ,
\label{eq:apxB.2}   \\
    &\phantom{}_{2} F_{1} 
    \biggl[ a+ \lambda , b- \lambda; c;\dfrac{1-z}{2} \biggr] \overset{\lambda \rightarrow \infty}{\sim}
    \dfrac{\Gamma(1-b + \lambda) \Gamma(c)}{\Gamma(\frac{1}{2}) \Gamma(c- b + \lambda)} 2^{a+b - 1} (1- e^{-\xi})^{-c+ \frac{1}{2}}(1+ e^{-\xi})^{c-a-b- \frac{1}{2}} \lambda^{-\frac{1}{2}} \notag \\
    &\times [e^{(\lambda - b)\xi} + e^{\pm \iu \pi (c- 1/2)}e^{-(\lambda +a) \xi}][1+ \mathcal{O}(|\lambda^{-1}|)] \,, \label{eq:apxB.3}
\end{align}      
\end{widetext} 
where $\xi$ is defined by $e^{\xi} \equiv z+\sqrt{z^{2} -1}$, and the upper or lower sign is chosen according to whether $\mathrm{Im} (z) \gtrless 0 $. Equivalently, $e^{-\xi} = z- \sqrt{z^{2}-1}$. For real $z$ arguments, one may consider $z = a \pm \iu \epsilon$ and take $\epsilon \rightarrow 0$ that leaves a consistent result in the asymptotic expansion. 

\onecolumngrid
\subsection{Radius of convergence}
The asymptotic behavior of the odd-$l$ angular solution as $k \rightarrow  \infty$ is provided in Eq.~(B27) of Gralla. The corresponding even-$l$ angular solution is obtained by setting $a =  1/2$, $b = 0$, $c = 3/2$, $\lambda = k$, and $z = 1 - 2 \cos^{2}{\theta}$ in Eq.~\eqref{eq:apxB.3},
\begin{align}
    \theta_{2k}(\theta) & \overset{k \rightarrow \infty}{\sim} - \frac{1}{2 \sqrt{2 k} } \dfrac{\Gamma(k+1)}{\Gamma(k+ \frac{3}{2})} \dfrac{e^{\iu \theta} \sin{\theta}}{\sqrt{1- e^{2 \iu \theta}}} [(-e^{-2\iu \theta})^{k+ 1/2} - (-e^{-2\iu \theta})^{-k}] . \label{eq:apxB.4}
\end{align}
This asymptotic expression of $\theta_{2k}$ is real at all $\theta$, but it generally requires a careful cancellation of the complex phases. In addition, the asymptotic behavior of the even-$l$ inner radial solution~\eqref{eq:2.31} in the same limit can be obtained by setting $a=2$, $b=1$, $\lambda = 2k$, $c=3$, $z = 1- r/M$ in Eq.~\eqref{eq:apxB.3},
\begin{equation}
    R^{<}_{2k}(r)
    \overset{k \rightarrow \infty}{\sim} 
     \sqrt{\dfrac{8}{\pi k}} \frac{\Gamma(2k)\Gamma(2k+2)}{\Gamma(4k+1)}(2M)^{2k-1} r^{2}  \dfrac{(  e^{-2k \xi} + \iu e^{(2k+1)\xi} )}{(1-e^{\xi})^{2} \sqrt{e^{-2\xi} -1}} , \:\:\:\:\: e^{\xi} = 1- \dfrac{r- \sqrt{r(r-2M)}}{M} \, , \label{eq:apxB.5}
\end{equation}
where lower sign is chosen so that the real part of the radial function is always positive. Note that $0<|e^{\xi}|<1$ for $2M<r$; hence the imaginary part in Eq.~\eqref{eq:apxB.5} virtually vanishes as $k \rightarrow \infty$. If we rewrite the inner solution as
\begin{equation}\label{eq:apxB.6}
    \psi^{<}(r,\theta) = C \sum^{\infty}_{k=1} \, a^{<}_{k} , \:\:\:\:\: a^{<}_{k} \equiv \frac{1}{\sqrt{\pi}} \dfrac{\Gamma(k+\frac{1}{2})}{\Gamma(k+1)} \frac{R^{<}_{2k}(r)}{d^{2k}} \Theta_{2k}(\theta) ,
\end{equation}
the asymptotic behavior of the summation term, using Eqs.~\eqref{eq:apxB.4} and~\eqref{eq:apxB.5}, is
\begin{equation}\label{apxB.7}
    a^{<}_{k} \overset{k \rightarrow \infty}{\sim} 
     \dfrac{(-1)^{k}}{k \sqrt{\pi}} \dfrac{r^{2}}{2M} \dfrac{\Gamma(2k)}{ \, \Gamma(2k+ \frac{1}{2})} 
    \left( \dfrac{M}{2d} \right)^{2k}
    \sqrt{\dfrac{1-e^{2 \iu \theta}}{ e^{-2 \xi} - 1}} \dfrac{(  e^{-2k \xi} + \iu e^{(2k+1)\xi} )(e^{-(2k+1) \iu \theta} + \iu e^{2k \iu \theta})}{(1-e^{\xi})^{2}} \,.
\end{equation}
In general, the ratio $|a^{<}_{k+1}/a^{<}_{k}|$ wildly oscillates, and thus does not approach a smooth limit to perform the ratio test. But evaluating the summation term at $\theta = \pi /4$ allows the sum to be separated into even and odd $k$ contributions as 
\begin{equation}    \sum^{\infty}_{k=1}a^{<}_{k} = \sum^{\infty}_{m =1} a^{<}_{2m+1} + \sum^{\infty}_{n = 1} a^{<}_{2n} \, ,
\end{equation}
due to the angular part
\begin{equation}
    \Theta_{2k}(\theta= \pi/4) \overset{k \rightarrow \infty}{\sim} \frac{1}{4\sqrt{k}} \dfrac{\Gamma(k+1)}{\Gamma(k+ \frac{3}{2})} \left[ \sqrt{\sqrt{2}-1} \cos{\left( \frac{k \pi}{2}\right)} + \sqrt{\sqrt{2}+1} \sin{\left( \frac{k\pi}{2}\right)} \right] \, ,
\end{equation}
from which we see that $|a^{<}_{k+2}/a^{<}_{k}|$ does not oscillate wildly. Then, we may instead perform the ratio test on each contribution separately. We find that both contributions have the same ratio of consecutive terms in the limit $k \rightarrow\infty$,
\begin{equation}
    L = \lim_{m \rightarrow \infty} \left| \dfrac{a^{<}_{2m +1}}{a^{<}_{2m-1}} \right| = \lim_{n \rightarrow \infty} \left| \dfrac{a^{<}_{2n+2}}{a^{<}_{2n}}  \right| = \left( \dfrac{M}{2d} \right)^{4} |e^{-4 \xi}| = \left( \dfrac{M}{2d} \right)^{4}  \left[ \dfrac{r+ \sqrt{r(r-2M)}}{M} -1 \right]^{4} \,,
\end{equation}
Thus, by demanding that $L =1$, we find
\begin{equation}
    d = \dfrac{r - M + \sqrt{r(r-2M)}}{2} \, ,
\end{equation}
which gives the radius of convergence in Eq.~\eqref{eq:4.59}, such that $L < 1$ for every $2M<r < d_{\circ}$. Alternatively, although all $a_k$'s are zero at $\theta = \pi/2$, we may first take the limit $\theta \rightarrow \pi/2$ of the ratio $a_{k+1}/a_{k}$ and then perform the ratio test. This procedure yields the same radius of convergence.

For completeness, we also show that the outer solution has the same radius of convergence as the inner solution. The outer solution, on the other hand, has two summations, one of which arises from the $l =0$ stream equation. This summation is simply the last term evaluated at $r = d_{\circ}$ and $\theta = \pi /2$. Thus, establishing convergence of the last summation at $\theta = \pi/2$, where the ratio of the consecutive terms does not fluctuate wildly, is sufficient to prove that $r = d_{\circ}$ is the radius of convergence of the full solution.

Since the asymptotic behavior of the odd-$l$ angular solution is already known, we only need to show here the asymptotic expression for $R^{>}_{2k - 1}(r)$. Applying the Pfaff transformation 
\begin{equation}
    \phantom{}_{2}F_{1} \left[ a,b;c;z \right] = (1-z)^{-a} \phantom{}_{2}F_{1}\left[a,c-b;c;z/(z-1)\right]
\end{equation}
to the odd-$l$ case in Eq.~\eqref{eq:2.33} gives 
\begin{equation}
    R^{>}_{2k-1}(r) = r^{-(2k-1)} \left( 1- \dfrac{2M}{r}\right)^{-(2k+1)} \phantom{}_{2}F_{1} \left[ 2k+1,2k+1;4k;\dfrac{1}{1-r/2M}\right] \,.
\end{equation}
Then, with the choice of $a = 1$, $b  = 2$, $ c = 1$, $\lambda = 2k$, and $z =  r/M - 1$ in Eq.~\eqref{eq:apxB.2}, the asymptotic expression for the odd-$l$ outer solution is 
\begin{equation}
    R^{>}_{2k - 1}(r) \overset{k \rightarrow \infty}{\sim}  \sqrt{\dfrac{32\pi}{k}} \left( \dfrac{e^{-\xi}}{2M} \right)^{2k+1} \frac{\Gamma(4k)}{\Gamma(2k+1)\Gamma(2k-1)} \dfrac{r^{2}}{(1+e^{-\xi})^{2}\sqrt{1-e^{-2\xi}}} \, ,
\end{equation}
where this time $0< e^{\xi} < 1$ for $  2M < r$. Rewriting the summation in the last term of the outer solution as 
\begin{equation}
   \sum^{\infty}_{k = 1} \dfrac{\Gamma(k - \frac{1}{2})}{\Gamma(k+1)} d^{2k} R^{>}_{2k -1}(r)\Theta_{2k - 1}(\theta) =  \sum^{\infty}_{k = 1} a^{>}_{k}  \, ,
\end{equation}
the summation term has an asymptotic behavior of 
\begin{align}
    a^{>}_{k} &\overset{k \rightarrow \infty}{\sim} \frac{(-1)^{k}4\sqrt{\pi}}{k} \dfrac{\Gamma(4k)}{\Gamma(2k)\Gamma(2k+1)} 
    \left( \dfrac{ e^{-\xi} d}{2M} \right)^{2k+1}
    \frac{r^{2} [e^{-2k \iu \theta } + \iu e^{(2k-1)\iu \theta}]}{d(1+ e^{-\xi})^{2}} \sqrt{\dfrac{1-e^{2\iu \theta}}{1- e^{-2\xi}}} 
\end{align}
At $\theta = \pi/2$, we find 
\begin{equation}
    L = \lim_{k \rightarrow \infty} \left| \dfrac{a^{>}_{k+1}}{a^{>}_{k}} \right| = \left( \dfrac{2e^{-\xi}d}{M} \right)^{2} = \left( \dfrac{2d}{M} \right)^{2} \left[ \dfrac{r-\sqrt{r(r-2M)}}{M} -1\right]^{2} \,, \label{eq:apxB.18}
\end{equation}
which gives the same radius of convergence as in Eq.~(\ref{eq:4.59}), such that $L < 1$ for every $r > d_{\circ}$. In both solutions, we did not show convergence at exactly $r = d_{\circ}$ for all $\theta$, but this can be verified numerically.

\subsection{Normalization constant}

Now that we have already established the convergence of the solutions, we determine the normalization constant in the inner solution by enforcing continuity at the common radius of convergence. In principle, the normalization constant should be independent of $\theta$. Hence, we fix the angle at $\theta = \pi/4$, and obtain a rapidly convergent summation terms that allows us to arrive at an approximate expression of $C$. 

We rewrite the outer solution at $(r = d_{\circ}, \theta= \pi/4)$ as
\begin{align}
    \psi^{>}(r = d_{\circ}, \theta = \pi/4) = \sum^{\infty}_{m = 1} b^{\text{odd}}_{m} &+ \sum^{\infty}_{n =1} b^{\text{even}}_{n} \,,  \\
    b^{\text{odd}}_{m} = - \frac{1}{2\sqrt{\pi}} \dfrac{\Gamma(2m - \frac{3}{2})}{\Gamma(2m)} & d^{4m-2} R^{>}_{4m-3}(d_{\circ})
    \left[\Theta_{4m-3}\left(\frac{\pi}{4} \right)  - 1+\frac{1}{\sqrt{2}}\right] \, , \\
    b^{\text{even}}_{n} = - \frac{1}{2\sqrt{\pi}} \dfrac{\Gamma(2n - \frac{1}{2})}{\Gamma(2n+1)}  & d^{4n} R^{>}_{4n-1} (d_{\circ})
    \left[\Theta_{4n-1}\left(\frac{\pi}{4} \right)  - 1+\frac{1}{\sqrt{2}}\right] \, .
\end{align}
At $r = d_{\circ}$, we have the identity $e^{-\xi} = \dfrac{M}{2b}$. Then, the asymptotic behavior of the radial part is
\begin{equation}
    R^{>}_{2k -1}(d_{\circ}) \overset{k \rightarrow}{\sim} 32\sqrt{\dfrac{2\pi}{k}} \dfrac{\Gamma(4k)}{\Gamma(2k+1)\Gamma(2k-1)}
    \left(\dfrac{1}{4d} \right)^{2k+1} 
    \dfrac{d^{2}_{\circ}d^{3}}{(2d+M)^{2}\sqrt{4d^{2}-M^{2}}} \,. \label{eq:apxB.21}
\end{equation}
Meanwhile, at $\theta = \pi/4$, the angular harmonics have the asymptotic form
\begin{equation}
    \Theta_{2k-1}\left(\theta = \frac{\pi}{4}\right) \overset{k \rightarrow}{\sim}
    \dfrac{(-1)^{k}}{2\sqrt{k}} \dfrac{\Gamma(k+1)}{\Gamma(k+ \frac{1}{2})} \left[ \sqrt{\sqrt{2}+1} \cos{\left( \frac{k \pi}{2} \right)} - \sqrt{\sqrt{2}-1} \sin{\left( \frac{k \pi}{2} \right)} \right] .
    \label{eq:apxB.22}
\end{equation}
Using the Stirling's approximation $\Gamma(z) \overset{z \rightarrow \infty}{\sim} \sqrt{2\pi} z^{z-1/2}e^{-z}$ in Eqs.~\eqref{eq:apxB.21}~and~\eqref{eq:apxB.22}, the odd and even summation terms exhibit
\begin{equation}
    b^{\text{odd}}_{m} \overset{m \rightarrow \infty}{\sim} c^{\text{odd}}_{m} \equiv \left[ \dfrac{(-1)^{m} \sqrt{2(\sqrt{2}-1)} -1+\sqrt{2}}{m^{3/2} \sqrt{\pi}} \right] \chi  , \:\:\:
    b^{\text{even}}_{n}  \overset{n \rightarrow \infty}{\sim} c^{\text{even}}_{n} \equiv \left[ \dfrac{2\sqrt{2} -2 - (-1)^{n} \sqrt{2(\sqrt{2}+1)}}{2n^{3/2} \sqrt{\pi}} \right] \chi\,,
\end{equation}
where we define $\chi$ as 
\begin{equation}
    \chi \equiv \dfrac{d^{2}_{\circ} d^{2}}{(2d+M)^{2} \sqrt{4d^{2} -M^{2}}}  .
\end{equation}
From these asymptotic behaviors, the slow convergence of the series can be attributed to the weak $k^{-3/2}$ decay together with the alternating nature of the terms. To obtain an approximate value for the outer solution, we sum the asymptotic contributions from $m,n =2$ to infinity, while retaining the exact first term of each contribution. Since the infinite sum of $k^{-n}$ is the Riemann Zeta function $\zeta(n)$, and the infinite sum of $(-1)^{k-1}/k^{s}$ is the Dirichlet zeta function $\eta(s)= (1-2^{1-s}) \zeta(s)$, we obtain
\begin{align}
    \sum^{\infty}_{m = 2} c^{\text{odd}}_{m} &= \frac{\alpha^{2}}{\sqrt{\pi}} \left[ \frac{\sqrt{2}}{\alpha}-1+(1-\alpha)\zeta\!\left(\dfrac{3}{2}\right) \right]
    \chi \, , \:\:\:\:
    \sum^{\infty}_{n = 2} c^{\text{even}}_{n} = 
    \frac{\alpha^{2}}{2\sqrt{\pi}}
\left[
-2-\frac{\sqrt{2}}{\alpha}
+\left(2+ \alpha \right)\zeta\!\left(\dfrac{3}{2}\right)
\right] \chi,
\end{align}
where $\alpha \equiv \sqrt{\sqrt{2}-1}$, and their sum is 
\begin{align}
   \sum^{\infty}_{m = 2} c^{\text{odd}}_{m} +  \sum^{\infty}_{n = 2} c^{\text{even}}_{n} &=\frac{\alpha}{2\sqrt{\pi}} \left[   \sqrt{2}  - 4\alpha -(\alpha - 4) \zeta\left( \dfrac{3}{2} \right) \right] \, \chi \,.
\end{align}
Hence, the outer solution has an approximate expression of 
\begin{align}
    \psi^{>}& 
    \approx \frac{\alpha}{2\sqrt{\pi}} \left[   \sqrt{2}  - 4\alpha -(\alpha - 4) \zeta\left( \dfrac{3}{2} \right) \right] \chi + \frac{d^{2}}{4096 M^{7}} 
    \biggl[ 768 (\sqrt{2}-1) M^{5} (M+ d_{\circ}) \notag \\
    &+ 70(2\sqrt{2} - 7) (2M^{3} +12 M^{2} d_{\circ} - 75 M d_{\circ}^{2} + 45 d^{3}_{\circ}) \, M d^{2} \notag \\
    &+ 3 d^{2}_{\circ} \log{\left( 1- \frac{2M}{d_{\circ}} \right)} \left[ 128(\sqrt{2}-1) M^{4} + 35 (2\sqrt{2} - 7) (24 M^{2} - 40 M d_{\circ} + 15 d^{2}_{\circ}) \, d^{2} \right] \biggr] \,. \label{eq:apxB.27}
\end{align}
We may apply the same procedure to the inner solution. However, we do not present the details here, and instead quote the resulting approximate expression,
\begin{equation}
    \psi^{<} \approx C \alpha \sqrt{\frac{2}{\pi}} \left[ - \sqrt{2} + \alpha^{2} \zeta\left( \frac{3}{2}\right) \right] \, \chi + \dfrac{d^{2}_\circ}{896 \sqrt{2} d^{4}} \left[ 40 M^{3} - 120 M^{2} d_{\circ} + 105 M d_{\circ}^{2} + 112 (2d_{\circ} - 3M) d^{2} - 28 d^{3}_{\circ} \right]  \,.
    \label{eq:apxB.28}
\end{equation}
The expression for $C$ is then obtained by equating Eqs.~\eqref{eq:apxB.27}~and~\eqref{eq:apxB.28}. In the limit $d_{\circ} \gg M$ or $M \rightarrow 0$, $C$ should approach unity, and our truncated expression gives $C \approx 0.964$.

\section{Numerical approximation of \texorpdfstring{$\theta_{\circ}$}{theta} and its robustness}\label{apx.C}

\begin{figure*}[ht!]
    \centering
    \begin{minipage}{0.45\linewidth}
        \centering
        \includegraphics[width=\linewidth]{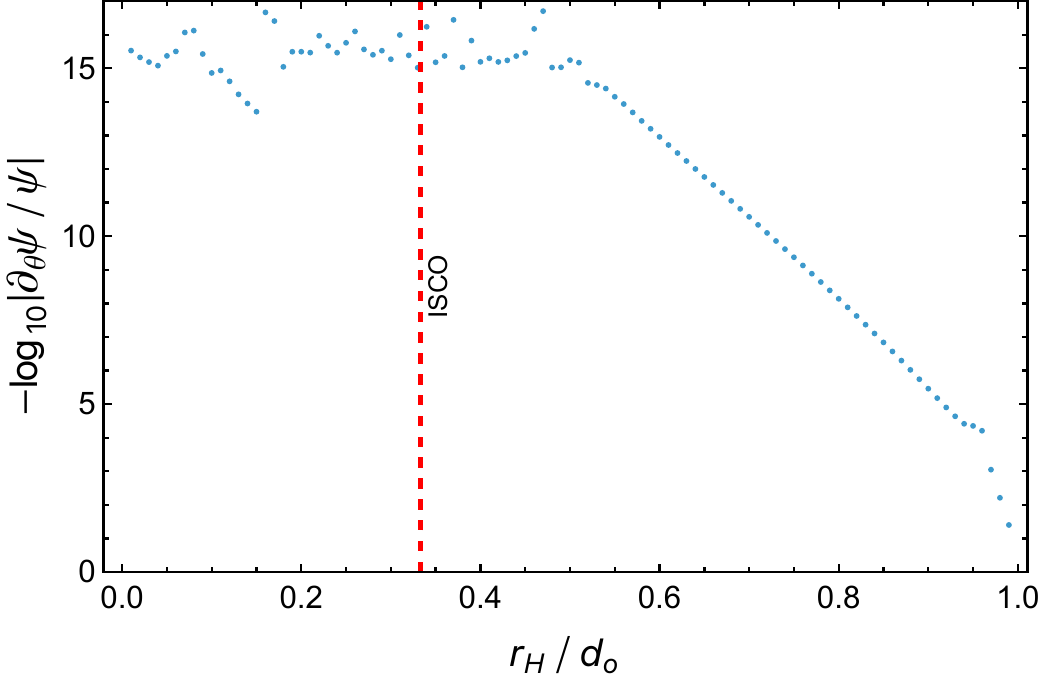}
        \caption{Robustness of our numerical approximation: $r_{H}/d_{\circ}$ vs. $- \log_{10}{|\partial_{\theta}\psi /\psi|}$ for some data points.  The red dashed line indicates $d_{\circ}=r_{\text{ISCO}}$.}
        \label{fig:4}
    \end{minipage}\hfill
    \begin{minipage}{0.45\linewidth}
        \centering
        \includegraphics[width=0.98\linewidth]{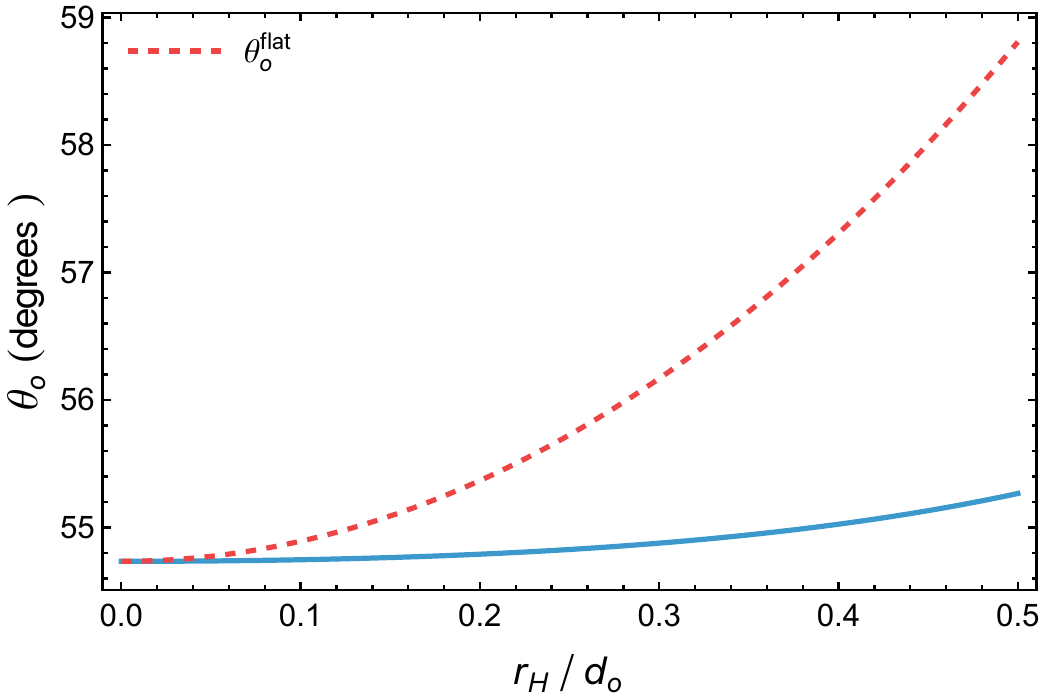}
        \caption{The values of $\theta_{\circ}$ for $0 \leq r_{H} / d_{\circ} \leq  0.5 $ using our numerical approximation (solid blue line) and exact expression~\eqref{eq:4.58} in flat spacetime (dashed red line).}
        \label{fig:5}
    \end{minipage}
\end{figure*}

\twocolumngrid
The inner flux function and its derivatives converges slowly. Nevertheless, we can obtain a precise approximation to the angle $\theta_{\circ}$ by obtaining the root of few terms from the summation of $\partial_{\theta}\psi$ evaluated at $r = r_{H}$. In our discussion, we used the first ten terms of the summation to determine $\theta_{\circ}$ for a given $d_{\circ}$. We find that $\partial_{\theta}\psi$, evaluated at $\theta_{\circ}$ on the horizon, vanishes to machine precision using our numerical approximation for large values of $d_{\circ}$, and even for values of $d_{\circ}$ within the ISCO, as shown in Fig.~\ref{fig:4}. However, our numerical approximation becomes imprecise as $d_{\circ}$ approaches the horizon. Also, we plot our numerical approximation of $\theta_{\circ}$ and the exact values of $\theta^{\text{flat}}_{\circ}$ for $0 \leq r_{H} / d_{\circ} \leq  0.5 $ in Fig.~\ref{fig:5}. The angles agree in large-$d_{\circ}$ limit.
\onecolumngrid
\twocolumngrid

%

\end{document}